\documentclass[aps,prc,reprint,amsmath,nofootinbib,superscriptaddress]{revtex4-2}

\usepackage{amsmath}
\usepackage[margin=1in]{geometry}
\usepackage{graphicx}
\usepackage{hyperref}
\usepackage{xcolor}
\graphicspath{{fig/}}

\DeclareMathOperator{\arccosh}{arccosh}
\DeclareMathOperator{\arctanh}{arctanh}
\DeclareMathOperator{\sech}{sech}

\newcommand{\avgxloss}{\left<\gamma_c\right>}
\newcommand{\bNN}{b_\mathrm{NN}}
\newcommand{\cssq}{c_s^2}
\newcommand{\eps}{\varepsilon}
\newcommand{\epsfb}{\varepsilon_\mathrm{fb}}
\newcommand{\epsFO}{\varepsilon_\mathrm{FO}}
\newcommand{\epsfrag}[1]{\varepsilon_{\mathrm{frag},#1}}
\newcommand{\epsG}{\varepsilon_\mathrm{G}}
\newcommand{\epsIC}{\varepsilon_\mathrm{IC}}
\newcommand{\ET}{E_\mathrm{T}}
\newcommand{\etas}{\eta_s}
\newcommand{\etascm}{\eta_{s\mathrm{,cm}}}
\newcommand{\etasmax}{\eta_{s\mathrm{,max}}}
\newcommand{\etasgridmax}{\eta_{s\mathrm{,gridmax}}}
\newcommand{\FF}{\mathcal{F}}
\newcommand{\ffb}{f_\mathrm{fb}}
\newcommand{\ffrag}{f_\mathrm{frag}}
\newcommand{\flatness}{f}
\newcommand{\GeVperfmcu}{\mathrm{GeV}/\mathrm{fm}^3}
\newcommand{\GeVperfmsq}{\mathrm{GeV}/\mathrm{fm}^2}
\newcommand{\kTmin}{k_\mathrm{T,min}}
\newcommand{\LambdaQCD}{\Lambda_\mathrm{QCD}}
\newcommand{\LHe}{\varepsilon_\mathrm{LH}}
\newcommand{\LHpz}{p_{z,\mathrm{LH}}}
\newcommand{\LHvz}{v_{z,\mathrm{LH}}}
\newcommand{\LL}{\mathcal{L}}
\newcommand{\LLF}{\tilde{\LL}}
\newcommand{\logtime}{\xi}
\newcommand{\MUSIC}{{\sc music}}
\newcommand{\Nch}{N_\mathrm{ch}}
\newcommand{\netas}{n_{\etas}}
\newcommand{\normfb}{N_\mathrm{fb}}
\newcommand{\normfrag}{N_\mathrm{frag}}
\newcommand{\Npart}{N_\mathrm{part}}
\newcommand{\Nscale}{N_\mathrm{scale}}
\newcommand{\ntau}{n_\tau}
\newcommand{\pdetas}{\partial_{\etas}}
\newcommand{\pdlogtime}{\partial_{\logtime}}
\newcommand{\pT}{p_\mathrm{T}}
\newcommand{\qty}[2]{#1~\text{#2}}
\newcommand{\rap}{y}
\newcommand{\rapbeam}{\rap_\mathrm{beam}}
\newcommand{\rFO}{r_\mathrm{FO}}
\newcommand{\sigmaG}{\sigma_\mathrm{G}}
\newcommand{\sigmainel}{\sigma_\mathrm{NN}^\mathrm{inel}}
\newcommand{\sNN}{s_\mathrm{NN}}
\newcommand{\sqrts}{\sqrt{\sNN}}
\newcommand{\tauFO}{\tau_\mathrm{FO}}
\newcommand{\tauIC}{\tau_\mathrm{IC}}
\newcommand{\term}[1]{{\textbf{#1}}}
\newcommand{\TFO}{T_\mathrm{FO}}
\newcommand{\Tmunu}{T^{\mu\nu}}
\newcommand{\trento}{T\raisebox{-0.5ex}{R}ENTo}
\newcommand{\trentoDDD}{T\raisebox{-0.5ex}{R}ENTo\nobreakdash-3D}
\newcommand{\trentoDDDalt}{T\raisebox{-0.5ex}{R}ENTo-3D}
\newcommand{\uT}{u_\mathrm{T}}
\newcommand{\vT}{v_\mathrm{T}}
\newcommand{\xT}{x_\perp}
\newcommand{\xTvec}{\vec{x}_\perp}

\begin{document}

\title{Bayesian parameter estimation with a new three-dimensional initial-conditions model for ultrarelativistic heavy-ion collisions}

\author{Derek Soeder}
\affiliation{Duke University, Durham, NC  27708-0305}
\author{Weiyao Ke}
\affiliation{Los Alamos National Laboratory, Los Alamos, NM  87545}
\author{J.-F. Paquet}
\affiliation{Vanderbilt University, Nashville, TN  37235-1807}
\author{Steffen A. Bass}
\affiliation{Duke University, Durham, NC  27708-0305}

\date{\today}

\begin{abstract}
We extend the well-studied midrapidity \trento{} initial-conditions model to three dimensions, thus facilitating (3+1)D modeling and analysis of ultrarelativistic heavy-ion collisions at RHIC and LHC energies.  \trentoDDD{} is a fast, parametric model of the 3D initial-state geometry, capable of providing initial conditions for (3+1)D models of quark--gluon plasma formation and evolution.  It builds on \trento{}'s success at modeling the initial nuclear participant thicknesses, longitudinally extending the initial deposition to form a central fireball near midrapidity and two fragmentation regions at forward and backward rapidities.  We validate the new model through a large-scale Bayesian calibration, utilizing as observables the rapidity distributions of charged hadrons.  For computational efficiency the present effort employs a (1+1)D linearized approximation of ideal hydrodynamics as a stand-in for quark--gluon plasma dynamics.  This calibration serves as model validation and paves the way for utilizing \trentoDDD{} as an initial-conditions model for state-of-the-art simulation incorporating (3+1)D relativistic viscous hydrodynamics.
\end{abstract}

\maketitle

\section{Introduction}

The determination of the properties of deconfined QCD matter, the quark--gluon plasma (QGP), has mostly relied on midrapidity heavy-ion collision data from symmetric collision systems at the Relativistic Heavy-Ion Collider (RHIC) and the Large Hadron Collider (LHC)~\cite{PHENIX:2003iij,PHENIX:2003qra,STAR:2008med,ALICE:2013mez}.  This was due to the collider architecture of these accelerator facilities and the corresponding design of their major detector systems (STAR and PHENIX at RHIC and ALICE at the LHC).  This undertaking has benefited from the sensitivity of certain observables measurable at midrapidity, such as the anisotropic flow coefficients $v_n$, to the QGP epoch of the collision \cite{Romatschke:2007mq,Song:2010mg,Gale:2012rq,Novak:2013bqa,Niemi:2015qia,Bernhard:2016tnd,Bernhard:2019bmu}.

However, a wealth of \emph{rapidity-dependent} experimental data does exist, in terms of measurements by the smaller BRAHMS and PHOBOS experiments \cite{BRAHMS:2001llo,PHOBOS:2001zjw,BRAHMS:2004dwr,PHOBOS:2004vcu}, in data from asymmetric collision systems \cite{STAR:2003pjh,PHENIX:2003qdw,BRAHMS:2004xry,PHOBOS:2010eyu,ALICE:2012eyl,ALICE:2013wgn}, and via the developing increases in rapidity coverage of the STAR, PHENIX, and ALICE experiments.  This type of data is expected to yield additional constraints on properties of the QGP and is crucial for improving our understanding of the initial-state and pre-equilibrium epochs of ultrarelativistic heavy-ion collisions. On the theory side, the availability of large time allocations on modern supercomputers has made it increasingly practical to conduct fully-(3+1)D simulations and reap the attendant benefits---particularly in the study of very asymmetric collisions systems, where modeling longitudinal structure is crucial.

Studies that model the QGP as an evolving medium do so using relativistic hydrodynamics, and therefore require an initial hydrodynamical state.  Typically this amounts to specification of the contents of the energy--momentum tensor $\Tmunu$  at some initial proper time $\tau_0$, by which time the system is assumed to be sufficiently locally equilibrated to permit hydrodynamic modeling.  The best choice of effective model to generate a representative sampling of initial conditions is not a settled matter, neither in 2D nor in 3D.  For the former, the Duke group has previously developed the \trento{} model~\cite{Moreland:2014oya}, a now well-established adaptation of the Monte Carlo Glauber and wounded-nucleon families of parametric models for the pre-hydrodynamic stage.  In (3+1)D, meanwhile, a variety of models has been considered (e.g. \cite{Bolz:1992nt,Geiger:1992ac,Hirano:2001eu,McDonald:2020oyf,Shen:2017bsr}, and an earlier 3D \trento{} formulation~\cite{Ke:2016jrd}), yet none has become an enduring basis for (3+1)D studies, and there is no consensus on the ``best'' principles for constructing such a model.

In this paper we present \trentoDDD{}, a new parametric model for fully 3D initial conditions.  \trentoDDD{} builds on \trento{}'s success at modeling nucleonic (and subnucleonic~\cite{Moreland:2018gsh}) configuration and participation and the resulting nuclear thicknesses, extending the scope of the model to a central fireball near midrapidity and two fragmentation regions at forward and backward rapidities.  In the next section we discuss the \trentoDDD{} model in detail; we then describe a separate, simplified evolution model used to connect the 3D initial conditions to final-state observables, and we verify the resulting model chain via comparison to simulated data.  Ultimately we demonstrate \trentoDDD{}'s flexibility through a Bayesian calibration to experimental data, and we conclude with consideration of future work to more fully calibrate the model.

\section{The \protect\trentoDDDalt{} Model}

\begin{figure}
  \includegraphics[width=0.9 \columnwidth]{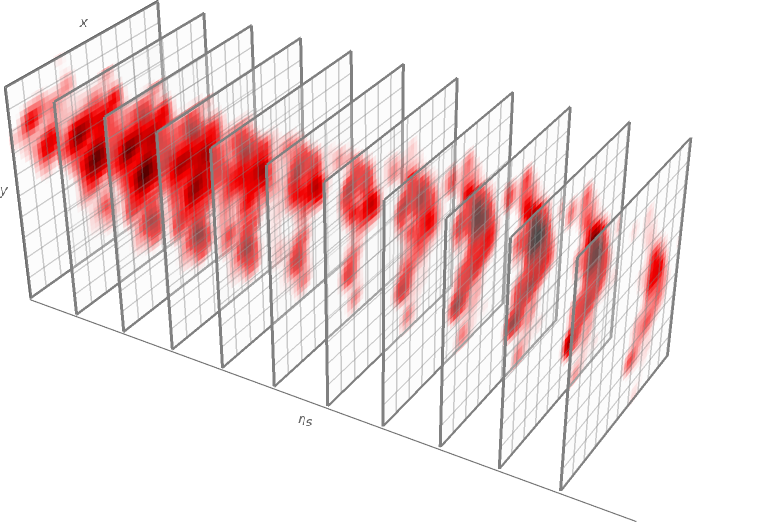}
  \caption
  {
    \label{fig:IC3D}
    Rendering of the initial conditions produced for a random Pb--Pb collision at $\sqrts = \qty{5.02}{TeV}$ and discretized onto a 3D grid.  For visualization the energy density (indicated by color and opacity) is flattened in the $\etas$ direction for each rapidity slice.
  }
\end{figure}

The starting point of the \trentoDDD{} model is to sample, for two colliding nuclei, the respective distributions of nuclear matter and the impact parameter that separates them.  Pairs of opposing nucleons are then tested for interaction to obtain the nuclear thicknesses that will determine the initial conditions.  This procedure is inherited from the 2D or boost-invariant \trento{}, so we review the salient concepts here and refer readers seeking a more detailed treatment to Refs.~\cite{Moreland:2014oya,Moreland:2018gsh}.  We then discuss the novel 3D extension, the essence of which is to transform these thicknesses into a spacetime rapidity-dependent energy deposition.  An example of \trentoDDD{} initial conditions is illustrated in Fig.~\ref{fig:IC3D}.

\subsection{Review of (2D) \protect\trento{}}
\label{sec:2D}

Given a collision system involving two nuclei $A$ and $B$ and a nucleon--nucleon center-of-mass energy $\sqrts$, \trento{} first generates the configuration of each nucleus by sampling positions in $(x, y, z)$ space for its nucleons, typically according to a Woods--Saxon distribution.  \trento{} then approximates the nuclei as infinitesimally thin nuclear pancakes (i.e., 2D projections) in the transverse ($x$--$y$) plane, owing to the Lorentz contraction along the $z$ direction in the ultrarelativistic limit.  At the moment of collision ($t = 0$), the two nuclear pancakes instantaneously overlap in the transverse plane at $z = 0$, their centers of mass offset in the transverse plane by some impact parameter $b$.

Every $A$ nucleon--$B$ nucleon pair is stochastically checked for inelastic collision, with the interaction probability based on the transverse separation $\bNN$ and the inelastic cross section $\sigmainel$ (a free parameter in \trento{}, while computed in \trentoDDD{} according to Eq.~\eqref{eq:sigmainel}).  Any nucleon that interacts at all becomes a ``participant'' in the collision; the nucleons that do not participate in any interaction are ``spectators'' and are not considered further.

Each nucleon comprises $n_c \geq 1$ \term{constituents}, which are modeled as 2D Gaussians in the transverse plane, each of uniform width $v$ and each normalized to have total thickness $1/n_c$ (since thickness expresses \emph{nucleon} number density).  The sum over the constituents of all participants in nucleus~$X$ yields the participant thickness $T_X$, defined
\begin{eqnarray}
&&T_X(\xTvec) = \sum_{p \in \mathrm{particip.\{X\}}} \frac{1}{n_c} \times \nonumber \\
&&\phantom{T_X(\xTvec) = }\sum_{c \in p} \gamma_c \frac{e^{-|\xTvec-\vec{x}_p-\vec{s}_c|^2 / 2 v^2}}{2\pi v^2}\text{,} \label{eq:TX}
\end{eqnarray}
where each subnucleonic constituent is fluctuated by a random factor $\gamma_c$ to account for the broad multiplicity distribution observed in proton--proton collisions.  $\vec{x}_p$ is the transverse position of the nucleon (relative to a fixed origin, so implicitly accounting for impact parameter) and $\vec{s}_c$ is the constituent's transverse position in the nucleon.

Finally, the two thicknesses are combined to produce the initial conditions.  \trento{}'s approach is to compute a ``reduced thickness'' using the generalized mean with a reduced thickness parameter $p$:
\begin{equation}
T_R(\xTvec) = \left( \frac{T_A^p(\xTvec) + T_B^p(\xTvec)}{2} \right)^{1/p}\text{.} \label{eq:TR}
\end{equation}
In effect this allows continuously varying among the behaviors of many 2D initial-conditions models.  Up to a normalization factor, $T_R$ is the distribution at midrapidity constituting \trento{}'s initial conditions.

\subsection{3D Ansatz}
\label{sec:3Dansatz}

In 2D \trento{}, the physical quantity that the initial conditions represented---energy or entropy density, for instance---was not determined a priori.  A key principle of \trentoDDD{} is that the initial conditions should represent an energy deposition, specifically the Bjorken-frame energy density at $\tau = \tauIC$:\footnote{
  The quantity $\ET$ appearing in Eq.~\eqref{eq:epsICdET} should not be confused with the final-state transverse energy observable $\ET = E \sin(\theta) = E \sech(\eta)$ (where $\eta$ is \emph{pseudo}rapidity) often encountered in experimental contexts.
}
\begin{equation}
\epsIC(\xTvec, \etas; \tauIC) = \frac{d^3 \ET}{d^2 \xT ~ \tau ~ d\etas}(\xTvec, \etas; \tauIC)\text{.} \label{eq:epsICdET}
\end{equation}
Here, the time coordinate $\tau = \sqrt{t^2 - z^2}$ (in natural units) is effectively a longitudinal proper time, the longitudinal coordinate $\etas = \arctanh(z/t)$ is spacetime rapidity, and $t$ and $z$ are the lab-frame time and longitudinal position---``lab frame'' being shorthand for the nucleon--nucleon center-of-momentum frame.

Such a specification of the initial energy density exhibits an unfortunate time dependence via the factor of $\tau$ in the denominator of Eq.~\eqref{eq:epsICdET}, which is due to the change of coordinates from $(t, z)$ to $(\tau, \etas)$.  Naturally the 3D energy density (commonly having units $\GeVperfmcu$) diverges as $\tau \to 0^+$, since the thickness in $z$ simultaneously vanishes.  However the quantity
\begin{align}
\lim\limits_{\tauIC \to 0^+} \tauIC ~ \epsIC(\xTvec, \etas; \tauIC) \qquad \qquad \nonumber \\
\qquad \qquad = \frac{d^3 \ET}{d^2 \xT ~ d\etas}(\xTvec, \etas; \tau = 0^+)
\end{align}
is well defined, and it is this quantity (in units of $\GeVperfmsq$) that \trentoDDD{} generates as initial conditions.  Accordingly it should be understood that there is no time dependence in the derivation that follows.  When $\tauIC$ appears for notational reasons, it simply cancels the time dependence of the accompanying energy density.

Equation~\eqref{eq:epsICdET} expresses the energy density co-moving with the Bjorken frame, which in the lab frame has velocity $\tanh(\etas)$ along the beamline.  We assume that, at this earliest moment of the collision, the initial distribution of nuclear matter consists of massless partons that do not significantly interact among themselves during the time $\tauIC$ before a dynamical stage begins.  Since all such partons originate at $z = 0$ and do not undergo any initial-stage accelerations, their longitudinal velocities correspond exactly to their spacetime rapidities \emph{and} their momentum rapidities (denoted $\rap$) via $v_z = \tanh(\etas) = \tanh(\rap)$.  Hence, in the Bjorken frame, all of the local nuclear matter's energy is carried by the transverse degree of freedom---a fact we emphasize with the subscript of $\ET$.

Specifying this energy distribution is the central problem of 3D initial conditions.  The \trentoDDD{} ansatz posits, for the longitudinal structure at any given transverse position $\xTvec$, a Gaussian-like \term{central fireball} around midrapidity, flanked by two \term{fragmentation regions} motivated by the limiting fragmentation hypothesis~\cite{Benecke:1969sh}.  This form is akin to the three-source model for particle production~\cite{Wolschin:2006nv} and is well-motivated by experiment as a \emph{final-state} phenomenon~\cite{PHOBOS:2004zne,ALICE:2016fbt}.  We contend that, sufficiently parametrized, this same form can permit a useful description of the initial state.

\begin{figure}
  \includegraphics[width=0.9 \columnwidth]{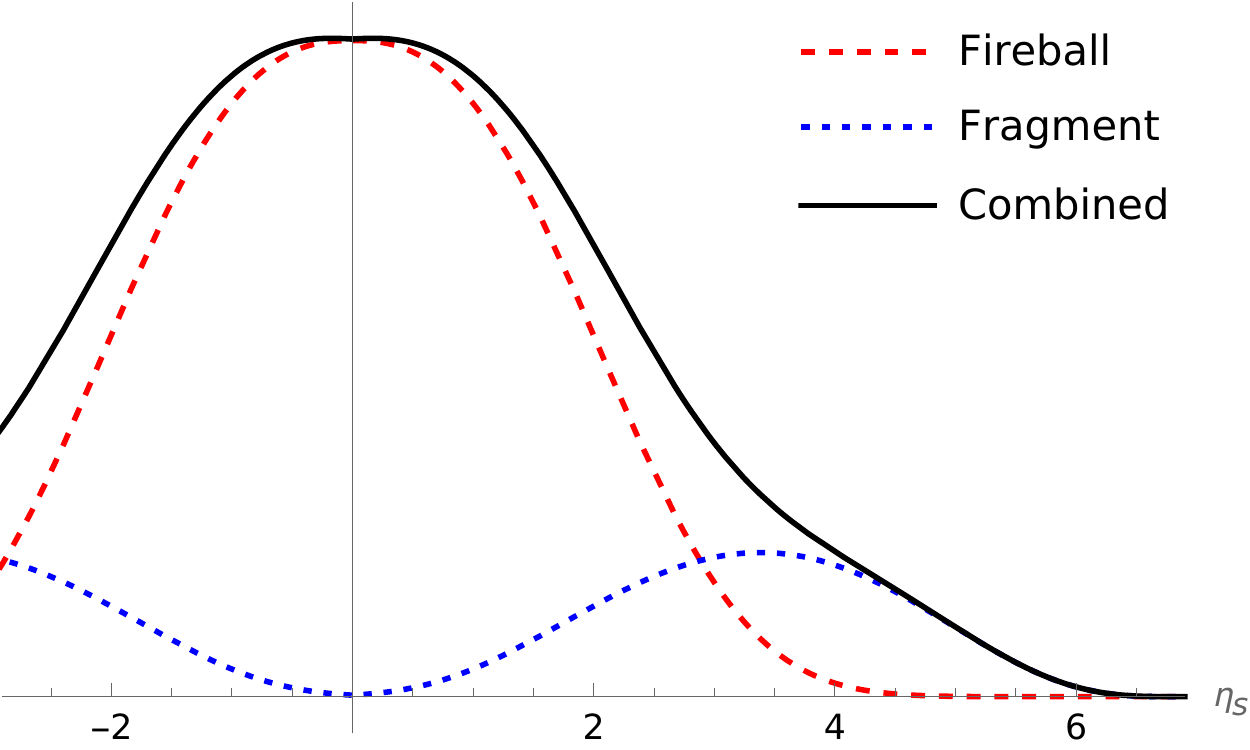}
  \caption
  {
    \label{fig:profiles}
    Illustration of the central fireball profile (red, dashed), the fragmentation profiles (blue, dotted), and their sum (black, solid), at $\sqrts = \qty{5.02}{TeV}$ with an arbitrary choice of parameter values.
  }
\end{figure}

In extending the \trento{} model to three dimensions, \trentoDDD{} assigns $T_X$ to be the \term{fireball thickness} and introduces a complementary \term{fragment thickness} $F_X$.  The energy deposited in the central fireball is proportional to the geometric mean of the fireball thicknesses, $\sqrt{T_A(\xTvec) ~ T_B(\xTvec)}$ (Eq.~\eqref{eq:TR} with $p = 0$), multiplied by a Gaussian-like envelope function, the \term{fireball profile function} (Fig.~\ref{fig:profiles}, red).

The choice to fix the value of $p$ marks a significant departure from the original \trento{} model, but it is compelled by energy--momentum conservation.  Specifically, the center-of-momentum-frame energy density of the combined fireball thicknesses, in the ultrarelativistic limit, is found to be proportional to $\sqrt{ \sNN ~ T_A(\xTvec) ~ T_B(\xTvec) }$~\cite{Shen:2020jwv}, thus requiring $p = 0$.  We also note that a large number of analyses \cite{Bernhard:2016tnd,Ke:2016jrd,Bernhard:2018hnz,Moreland:2019szz,Bernhard:2019bmu,Moreland:2018gsh,JETSCAPE:2020mzn,Giacalone:2020dln,Nijs:2020roc,Liyanage:2023nds,Giacalone:2023cet} have found that experimental data at ultrarelativistic energies is consistent with $p = 0$.  (Another analysis~\cite{Giacalone:2017uqx} is equivocal.)

As will be explained below, the fireball is not a single Gaussian-like shape symmetric about any one $\etas$; rather, it is a collection of longitudinally-extended contributions varying as a function of transverse position, with each contribution having the center of momentum of the two interacting patches of fireball thickness due to momentum conservation.  The patches of \emph{fragment} thickness, meanwhile, avoid the fireball, and each deposits its energy in its direction of travel such that on the $X$-going side, the energy deposited is $F_X$ times a \term{fragment profile function} (Fig.~\ref{fig:profiles}, blue).  At a high level, the entirety of the initial conditions is expressed as the sum
\begin{eqnarray}
\epsIC(\xTvec, \etas) &=& \epsfb(\xTvec, \etas) + \nonumber \\
&& \epsfrag{A}(\xTvec, \etas) + \epsfrag{B}(\xTvec, \etas) \text{,} \quad \label{eq:epsIC}
\end{eqnarray}
where the right-side terms respectively denote the energy density contributed by the central fireball, the $A$-going fragments, and the $B$-going fragments.  In subsequent sections we will describe how each contribution is computed, but first we discuss the model's approach to allocating energy among them.

\subsection{Energy Allocation}
\label{sec:fluct}

In \trentoDDD{}, unlike in the 2D \trento{} model, the entire energy--momentum balance of the collision system's participating matter must be accounted for, whether it ends up near midrapidity or in a fragmentation region.  Local conservation of energy and momentum is ensured by imposing the respective conditions
\begin{align}
\int \cosh(\etas) ~ d\etas ~ \tauIC ~ \epsIC(\xTvec, \etas) = \frac{\sqrts}{2} ~ \times \nonumber \\
\left( T_A(\xTvec) + F_A(\xTvec) + T_B(\xTvec) + F_B(\xTvec) \right)
\end{align}
and
\begin{align}
\int \sinh(\etas) ~ d\etas ~ \tauIC ~ \epsIC(\xTvec, \etas) = \sqrt{\frac{\sNN}{4} - m_p^2} ~ \times \nonumber \\
\left( T_A(\xTvec) + F_A(\xTvec) - T_B(\xTvec) - F_B(\xTvec) \right)\text{.}
\end{align}
Within $\epsIC$ (Eq.~\eqref{eq:epsIC}), however, the division of the $X$-going matter's energy between $\epsfb$ and $\epsfrag{X}$ is not constrained.  The \trentoDDD{} ansatz supposes a given $X$-going constituent, $c$, contributes a fraction $\gamma_c$ of its thickness to the central fireball, with $0 < \gamma_c < 1$, and the remaining fraction $1 - \gamma_c$ to the $X$-going fragmentation region.

The probability distribution of $\gamma_c$ is assumed to be a beta distribution:
\begin{equation}
\gamma_c \sim \mathrm{Beta}\left\lbrace \avgxloss \frac{1-k}{k}, \left(1 - \avgxloss\right) \frac{1-k}{k} \right\rbrace\text{.} \label{eq:betadistro}
\end{equation}
The parametrization of Eq.~\eqref{eq:betadistro} is chosen so that the mean of the beta distribution is $\avgxloss$, ensuring that on average an $X$-going participant thickness will be allocated between the fireball and $X$-going fragments with the expected proportions ($\avgxloss$ and $\left( 1 - \avgxloss \right)$ respectively).  In the present parametrization, $\avgxloss$ is a function of the energy content of the central fireball, which depends on $\sqrts$ and other model parameters as described in the next section.  An alternative to be considered in future work would make $\avgxloss$ a free parameter and compute the energy content of the fireball accordingly.

\begin{figure}
  \includegraphics[width=0.9 \columnwidth]{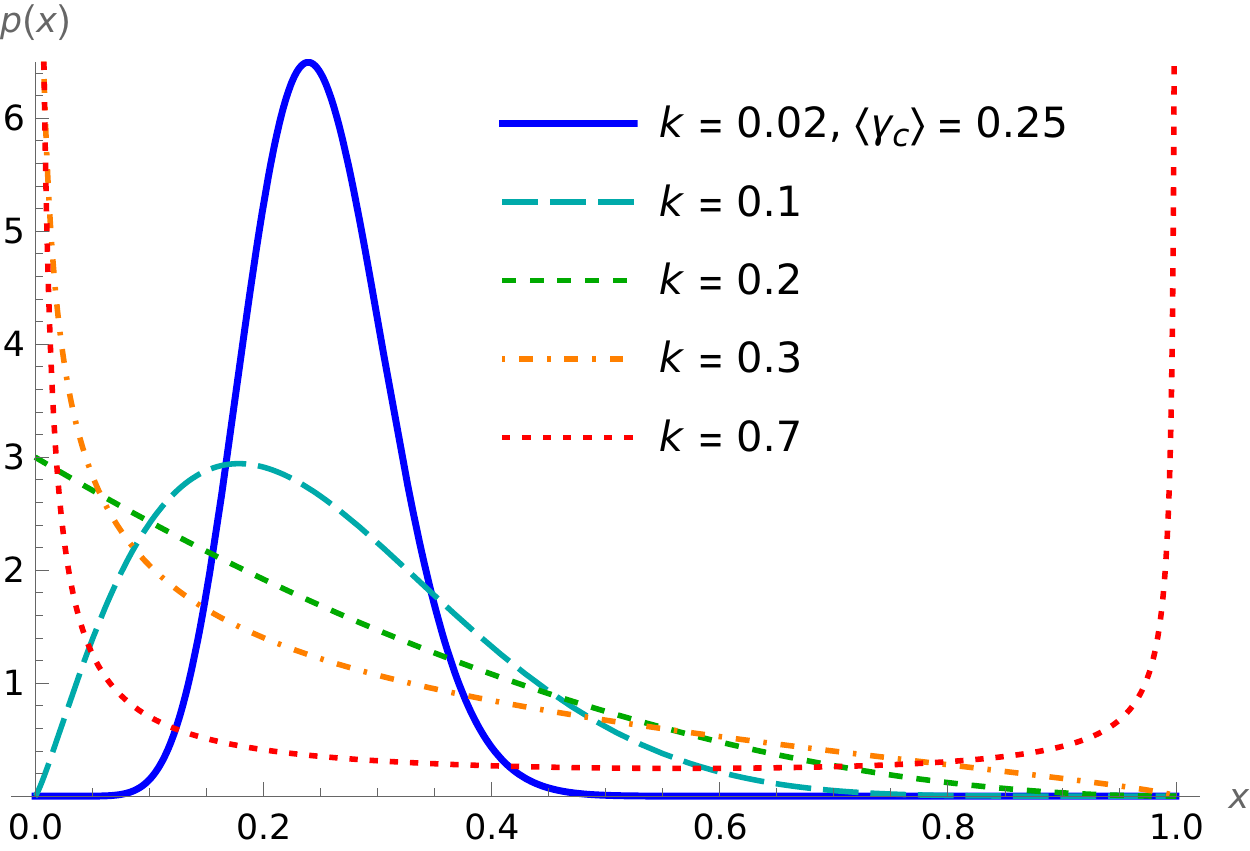} \\
  \includegraphics[width=0.9 \columnwidth]{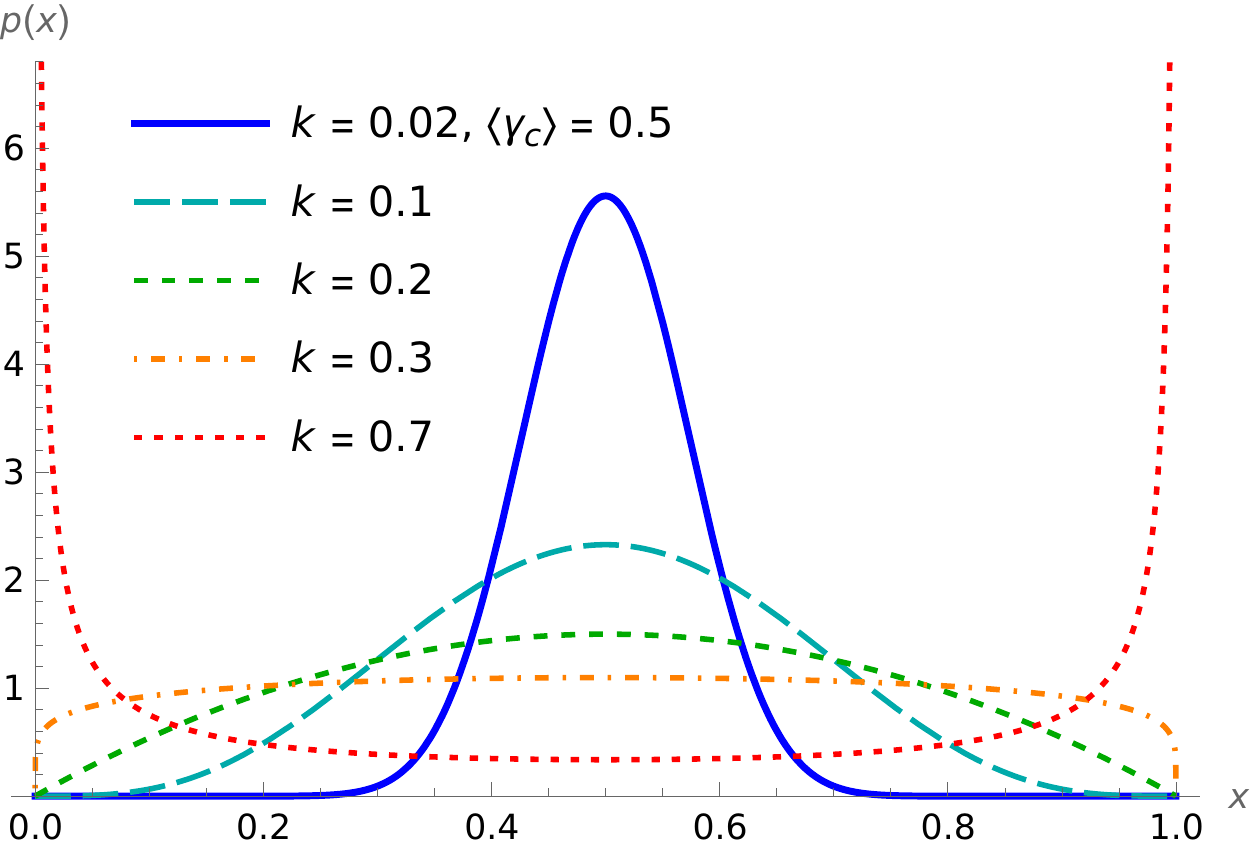}
  \caption
  {
    \label{fig:fluctuation}
    Distributions of the fireball thickness fraction $\gamma_c$ for two values of $\avgxloss$ (top, bottom) and five values of the fluctuation parameter $k$.
  }
\end{figure}

The \term{fluctuation parameter} $k \in (0,1)$ appearing in Eq.~\eqref{eq:betadistro} governs how the distribution is shaped about the given mean.  Typically, small $k$ favors $\gamma_c \sim \avgxloss$, moderate $k$ produces a broader distribution, and large $k$ skews the distribution toward extreme values such that any given constituent is likely to contribute mostly to the fireball or mostly to fragments, but is unlikely to be divided between them.  Some examples of the distribution~\eqref{eq:betadistro} are plotted in Fig.~\ref{fig:fluctuation} for various values of $k$ and $\avgxloss \leq 0.5$ (of which the $\avgxloss > 0.5$ distributions are mirror images).

For each nucleus~$X$, then, \trentoDDD{} samples $\gamma_c$ according to Eq.~\eqref{eq:betadistro} for each constituent of each participant and subsequently computes the two thickness functions, $T_X(\xTvec)$ for the central energy deposition as in Eq.~\eqref{eq:TX}, and $F_X(\xTvec)$ for the $X$-going fragments complementarily:
\begin{eqnarray}
&&F_X(\xTvec) = \sum_{p \in \mathrm{particip.\{X\}}} \frac{1}{n_c} \times \nonumber \\
&&\phantom{F_X(\xTvec) = }\sum_{c \in p} \left(1 - \gamma_c\right) \frac{e^{-|\xTvec-\vec{x}_p-\vec{s}_c|^2 / 2 v^2}}{2\pi v^2}\text{.}
\end{eqnarray}

As an aside, we remark that the original \trento{} model samples each constituent's contribution to $T_X$ from a gamma distribution, meaning $\gamma_c \in (0,\infty)$ in 2D \trento{}'s realization of Eq.~\eqref{eq:TX}.  Since \trentoDDD{} has both $T_X$ and $F_X$, $\gamma_c$ must now be interpreted as a fraction, a requirement that the beta distribution---with support over $(0,1)$---satisfies.

With the participant thickness thus decomposed, we can rewrite Eq.~\eqref{eq:epsIC} as
\begin{eqnarray}
&& \epsIC(\xTvec, \etas) = \epsfb(\xTvec, \etas; T_A, T_B) + \qquad \nonumber \\
&& \qquad \epsfrag{A}(\xTvec, \etas; F_A) + \epsfrag{B}(\xTvec, \etas; F_B)\text{.} \quad
\end{eqnarray}
The following two sections detail how $\epsfb$ and $\epsfrag{X}$ are computed from the relevant thicknesses.

\subsection{Central Fireball}
\label{sec:fb}

At the transverse position $\xTvec$, the central fireball carries a center-of-mass energy density $\sqrt{\sNN ~ T_A(\xTvec) ~ T_B(\xTvec)}$ with a center-of-mass rapidity
\begin{align}
\etascm(\xTvec) = \arctanh\Bigg( & \sqrt{1 - \frac{4 m_p^2}{\sNN}} ~ \times \nonumber \\
& \frac{T_A(\xTvec) - T_B(\xTvec)}{T_A(\xTvec) + T_B(\xTvec)} \Bigg)\text{,}
\end{align}
essentially the momentum rapidity of the sum of the $A$-going and $B$-going 4-momenta.

The central fireball contribution to the initial energy deposition is modeled by
\begin{align}
\epsfb(\xTvec,\etas) = \normfb ~ \sqrt{T_A(\xTvec) ~ T_B(\xTvec)} ~ \times \nonumber \\
\ffb(\etas-\etascm(\xT))\text{,} \label{eq:epsfb}
\end{align}
where $\normfb$ is an overall normalization parameter to be described below.  The fireball profile function $\ffb(\xTvec, \etas)$ is defined
\begin{align}
\ffb(\etas) = \exp\left\lbrace -\left(\frac{\etas^2}{2\left(\etasmax - \nu\right)}\right)^{\flatness} \right\rbrace \nonumber \\
\times \left(1 - \left(\frac{\etas}{\etasmax}\right)^4\right)^4\text{,} \label{eq:ffb}
\end{align}
with $\flatness$ a \term{flatness parameter} that interpolates between a Gaussian at $\flatness = 1$ and a plateau at larger $\flatness$.\footnote{
  The presence of a broad central plateau is a popular assumption, although it has been argued to be inconsistent with observations~\cite{PHOBOS:2004zne}.  We utilize the $\flatness$ parameter to describe a deformed Gaussian leading to a slightly flatter-than-Gaussian distribution, not an extended flat plateau per se.
}
The form of the Gaussian width is inspired by the particle production distribution of Ref.~\cite{Landau:1953gs}, with an additional \term{narrowing parameter} $\nu$ introduced for tuning.  The quartic portion is a \term{tapering factor} that ensures the fireball profile vanishes at finite $\etas$.

\begin{figure}
  \includegraphics[width=0.9 \columnwidth]{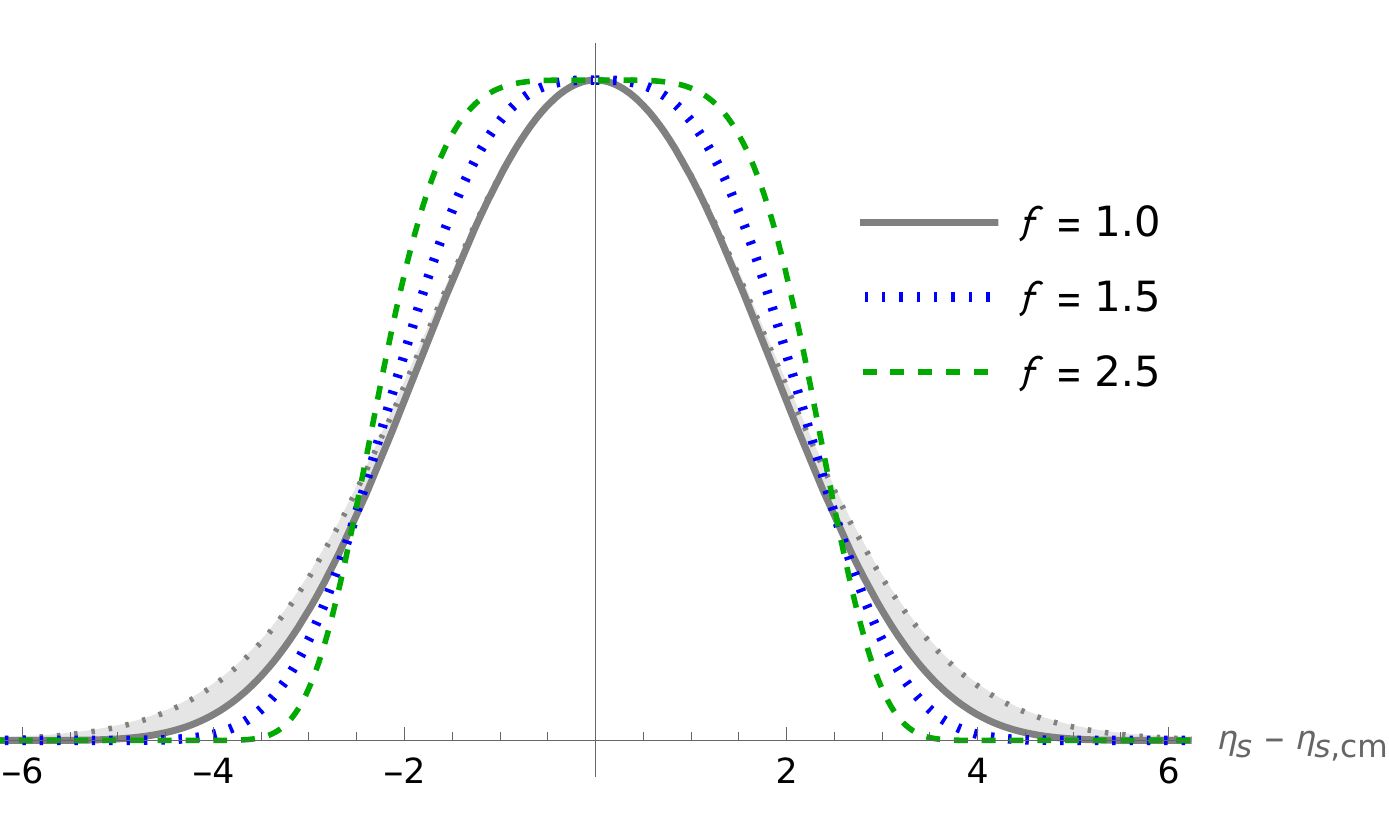}
  \caption
  {
    \label{fig:flatness}
    The central fireball profile at $\sqrts = \qty{200}{GeV}$, plotted with the flatness $\flatness$ varied and all other parameters fixed at arbitrary values.  The inner gray curve represents $\flatness = 1$, while the dotted outer edge of the gray band represents the same profile with the tapering factor removed---i.e., a Gaussian.  Also shown are the fireball profiles with $f = 1.5$ (blue, dotted) and $f = 2.5$ (green, dashed).
  }
\end{figure}

The quantity $\etasmax$ that appears in both the Gaussian-like factor and the tapering factor is defined
\begin{equation}
\etasmax = \arccosh\left(\frac{\sqrts / 2}{\kTmin}\right) \label{eq:etasmax}
\end{equation}
and essentially represents the maximum boost experienced by a massless parton of transverse momentum $\kTmin$ associated with the central fireball.  ($\kTmin$ will be discussed further in the next section.)  The effects of the flatness parameter and the tapering factor---both of which cause the fireball profile to deviate from a Gaussian---are illustrated in Fig.~\ref{fig:flatness}.

To complete the specification of Eq.~\eqref{eq:epsfb}, the energy scale of the central fireball must be defined via $\normfb$.  In an average nucleon--nucleon collision, we expect that each participant would contribute a fireball thickness $\avgxloss$, and that the pair would thus deposit energy $\avgxloss \sqrts$ into the central fireball.  Equating this with the energy capacity of the fireball profile, we obtain
\begin{equation}
\avgxloss \sqrts = \normfb \int_{-\etasmax}^{+\etasmax} \cosh(\etas) ~ d\etas ~ \ffb(\etas)\text{,}
\end{equation}
from which $\normfb$ can be computed if $\avgxloss$ is specified.  In the current parametrization, $\normfb$ is the model parameter and $\avgxloss$ is instead computed.

\subsection{Fragmentation Regions}
\label{sec:frag}

Just as the fireball thickness $T_X(\xTvec)$ is the density of $X$-going nuclear matter (in terms of nucleon number) contributing to the central fireball at a given $\xTvec$, so $F_X(\xTvec)$ expresses the density of matter contributing to the $X$-going fragmentation region.  As with the central fireball, the energy contributions of the fragmentation regions are modeled as the product of a thickness, a profile function, and a normalization.  The high-level expression is:
\begin{align}
\epsfrag{X}(\xTvec, \etas) = \frac{\kTmin}{\normfrag} ~ \times \qquad \nonumber \\
F_X(\xTvec) ~ \ffrag(e^{-\etasmax \pm \etas})\text{,}
\end{align}
where $F_X(\xTvec)$ is the fragment thickness mentioned above, $\pm$ is $(+)$ for $X = A$ and $(-)$ for $X = B$, and the other terms will be described below.

The limiting fragmentation hypothesis~\cite{Benecke:1969sh} contends that in the rest frame of a nucleus involved in a collision, a portion of the produced particles---referred to as ``fragments''---will conform to a momentum distribution that is independent of the collision energy~$E$ in the $E \to \infty$ limit.  These fragments can be understood as products of the break-up of the damaged nucleus, secondary to the collisions that form the central fireball.  One well-established consequence~\cite{Sahoo:2018osl} is that the rapidity spectrum of fragments approaches a limiting curve in the vicinity of $\rapbeam = \arccosh\left(\sqrts / 2 m_p\right)$; however, the limiting fragmentation hypothesis does not prescribe a shape for the trailing portion of the spectrum.  We assume a form for the fragment profile function motivated by parton distribution functions~\cite{Bass:2002vm,Gluck:1994uf}, specifically
\begin{equation}
\ffrag(x) = (-\ln x)^\alpha ~ x^{\beta + 1} ~ \exp\left(-\frac{2 \kTmin}{x \sqrts}\right)\text{,} \label{eq:ffrag}
\end{equation}
where the argument $x$ is analogous to the hadron momentum fraction and is computed
\begin{equation}
x = e^{-\etasmax + \lvert\etas\rvert}\text{,}
\end{equation}
in keeping with the customary relation for rapidity $\rap = \ln(1/x)$.

\begin{figure}
  \includegraphics[width=0.9 \columnwidth]{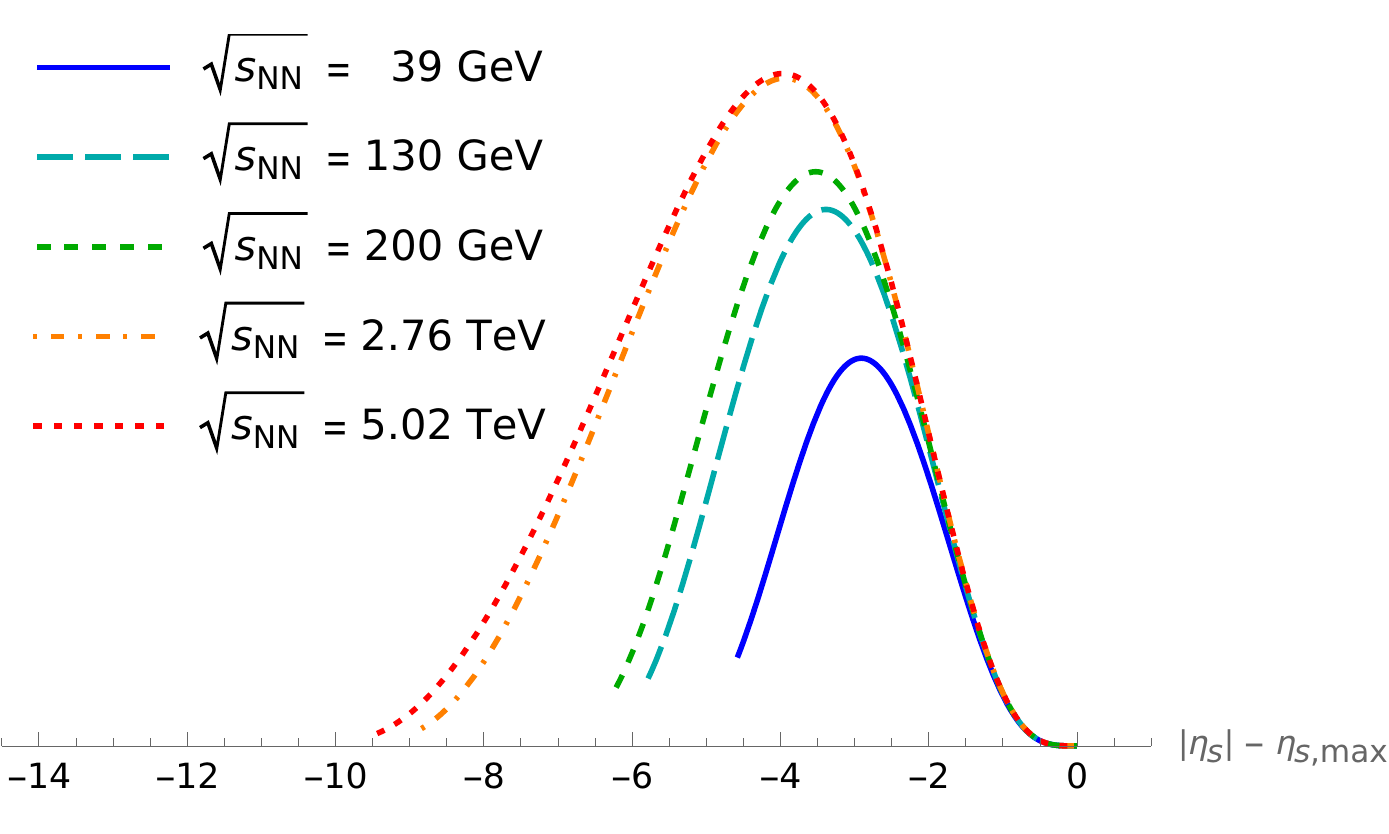}
  \caption
  {
    \label{fig:fragsqrts}
    Fragmentation profile~\eqref{eq:ffrag} plotted versus the spacetime rapidity offset from $\etasmax$ (Eq.~\eqref{eq:etasmax}) for various beam energies, with $\alpha$, $\beta$, and $\kTmin$ fixed to arbitrary values.  A common limiting-fragmentation curve is evident for all energies as $\lvert\etas\rvert$ approaches $\etasmax$.
  }
\end{figure}

As Fig.~\ref{fig:fragsqrts} illustrates, $\ffrag(e^{-\etasmax + \lvert\etas\rvert})$ shows a limiting fragmentation behavior in its leading edge as $\lvert\etas\rvert$ approaches $\etasmax$: in this region, the profiles converge to a single curve for all (ultrarelativistic) $\sqrts$.  The spacetime rapidity of this leading edge, $\etasmax$ (Eq.~\eqref{eq:etasmax}), does depend on beam energy; the important point is that the behavior in the vicinity of $\etasmax$ does not.

Overall we have introduced three parameters to describe the fragmentation regions: two fragment profile \term{shape parameters}, $\alpha$ and $\beta$, and the \term{transverse momentum scale} $\kTmin$, which also influences the central fireball. As an effective lower limit on the massless partons' transverse degree of freedom, $\kTmin$ entails a maximum spacetime rapidity and thereby ensures the longitudinal finitude of all components of the initial conditions:  It assigns a typical transverse momentum to the fragment partons, thus determining the limiting rapidity which is represented by $\etasmax$ and therefore by $x = 1$ as well.  $\kTmin$ also limits the extent of the central fireball (through the $\etasmax$-dependent tapering factor) and factors into its width (again via $\etasmax$), as discussed in Sec.~\ref{sec:fb}.

Completing Eq.~\eqref{eq:ffrag}, the normalization $\normfrag$ is computed
\begin{equation}
\normfrag = \int_{x_0}^1 dx ~ \ffrag(x)\text{,}
\end{equation}
with the lower limit of integration being the minimum (i.e. midrapidity) momentum fraction
\begin{align}
x_0 & = \exp\left( -\etasmax \right) \nonumber \\
& = \exp\left( -\arccosh\left(\frac{\sqrts / 2}{\kTmin}\right) \right) \nonumber \\
& \approx \frac{\kTmin}{\sqrts} \quad (\sqrts \gg \kTmin)\text{.}
\end{align}

Note that the fragment profile function, including $\normfrag$, is the same for both fragmentation regions.  It is the two fragment \emph{thicknesses} that will contribute differently to the two regions, based on the collision system and event-by-event fluctuations; the fragment \emph{profile} only specifies how those contributions will be distributed over $\etas$.

\subsection{Parameters}
\label{sec:params}

\begin{table*}
  \centering
  \begin{tabular}{|p{0.2 \textwidth}|c|c|p{0.6 \textwidth}|}
    \hline
    \textbf{Parameter} & \textbf{Symbol} & \textbf{Units} & \textbf{Description} \\
    \hline
    Constituent number & $n_c$ & & Number of constituents---subnucleonic degrees of freedom, or ``hotspots''---per nucleon (Sec.~\ref{sec:2D}). \\
    \hline
    Nucleon width & $w$ & fm & Width of the 2D-Gaussian thickness deposited by a wounded nucleon if $n_c = 1$, or width of the 2D normal distribution of constituent positions if $n_c > 1$. \\
    \hline
    Constituent width & $v$ & fm & Width of the 2D-Gaussian thickness deposited by a constituent, when $n_c > 1$ (Sec.~\ref{sec:2D}). \\
    \hline
    Form width & $u$ & fm & Effective nucleon width used to compute the nucleonic ($n_c = 1$) or ``partonic'' ($n_c > 1$) collision probability. \\
    \hline
    Fluctuation & $k$ & & Controls the parametrization of the distribution from which each participant or constituent draws the fraction of its thickness that contributes to the central fireball (Sec.~\ref{sec:fluct}). \\
    \hline
    Fireball normalization & $\normfb$ & GeV & Central fireball energy scale; conceptually similar to the normalization parameter in 2D \trento{} (Sec.~\ref{sec:fb}). \\
    \hline
    Flatness & $\flatness$ & & Exponent used to flatten the Gaussian-like central fireball profile (Sec.~\ref{sec:fb}). \\
    \hline
    Fragmentation \newline shape parameters & $\alpha$, $\beta$ & & Exponents in terms of which the fragmentation profile function is parametrized (Sec.~\ref{sec:frag}). \\
    \hline
    Transverse \newline momentum scale & $\kTmin$ & GeV & Typical transverse momentum of a parton; influences the shapes and extents in rapidity of the central fireball (Sec.~\ref{sec:fb}) and fragmentation regions (Sec.~\ref{sec:frag}). \\
    \hline
  \end{tabular}
  \caption
  {
    \label{tab:params}
    Summary of \trentoDDDalt{} model parameters.
  }
\end{table*}

\begin{figure*}
  \includegraphics[width=0.32 \textwidth]{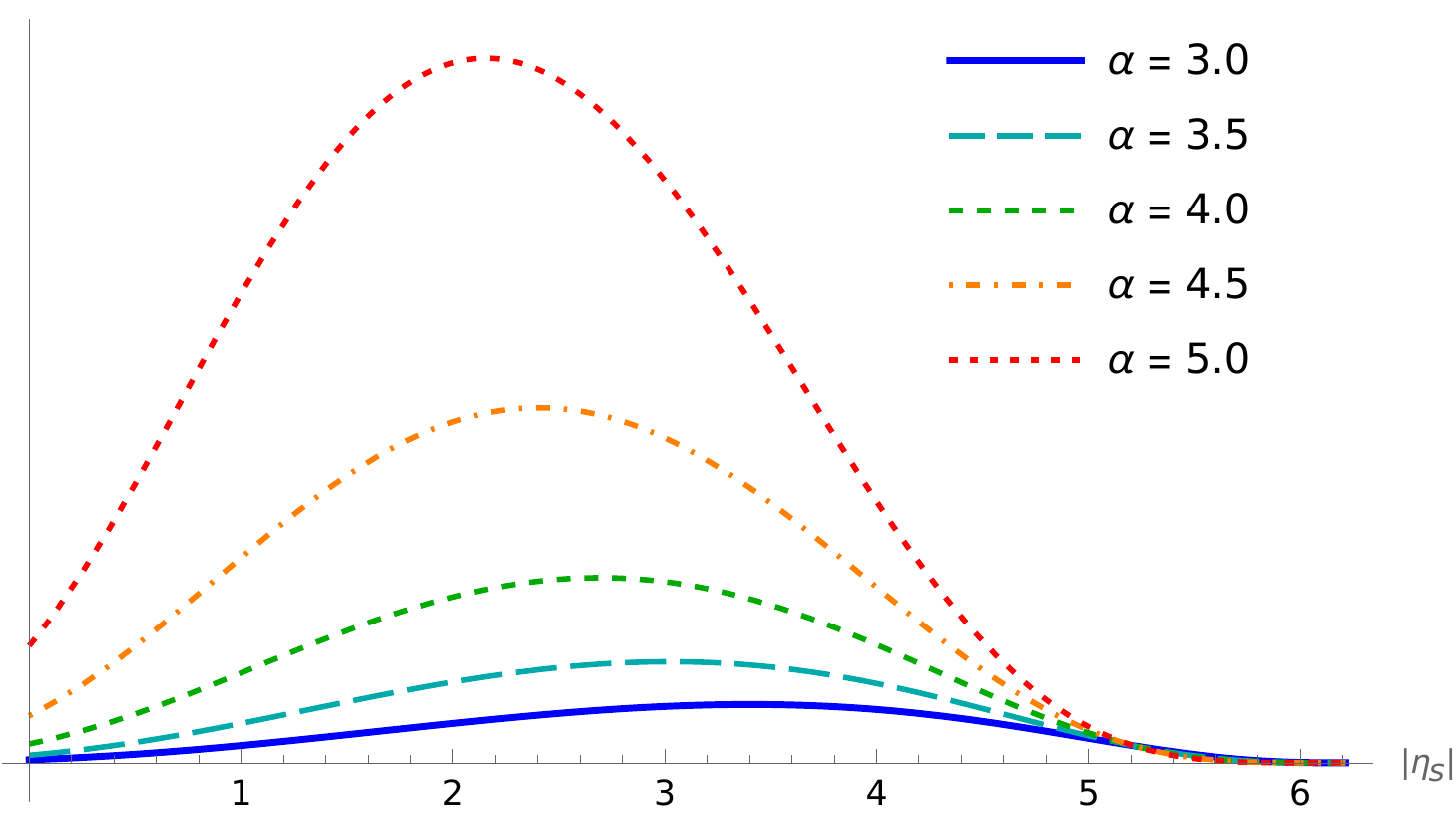} ~
  \includegraphics[width=0.32 \textwidth]{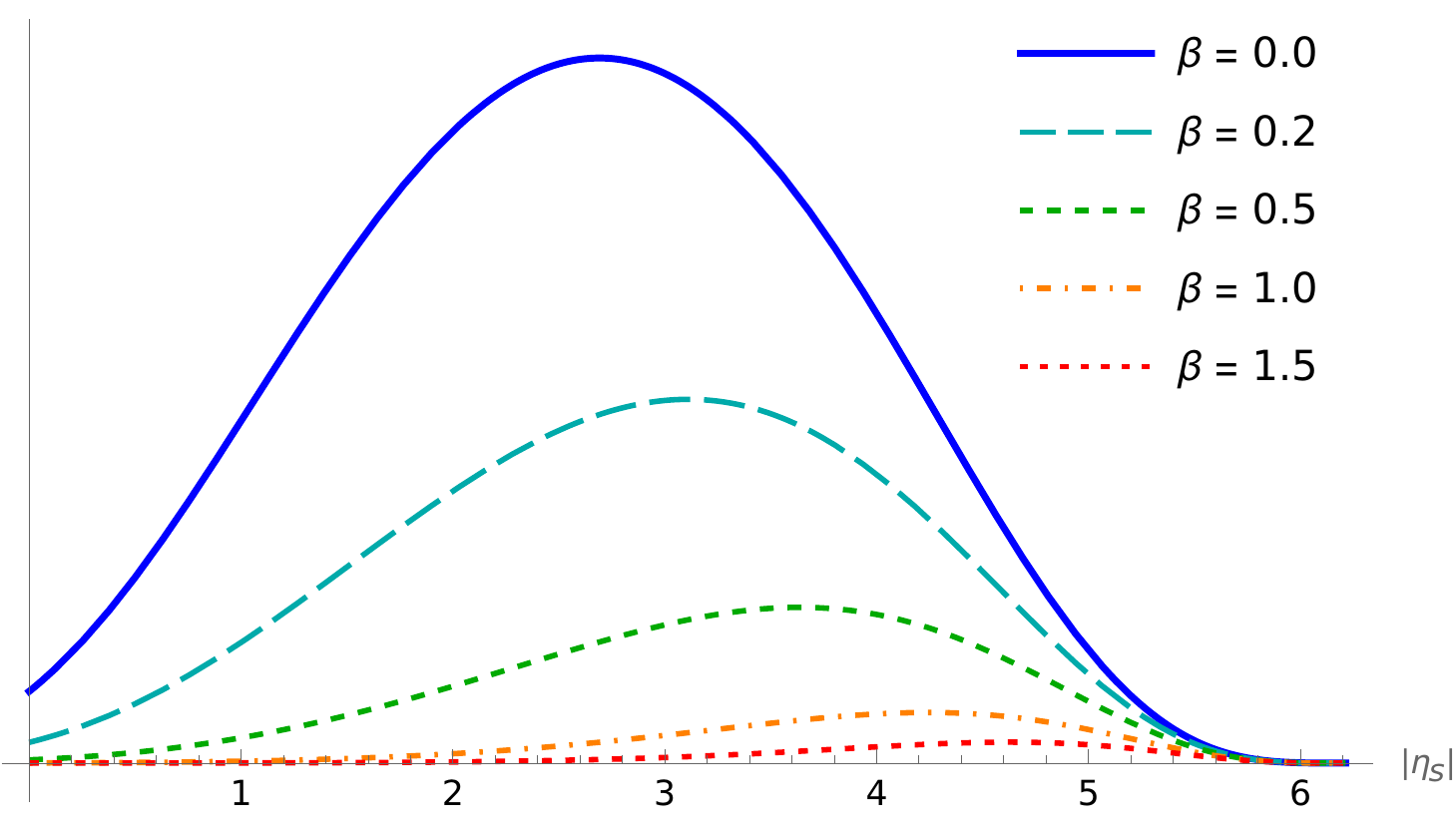} ~
  \includegraphics[width=0.32 \textwidth]{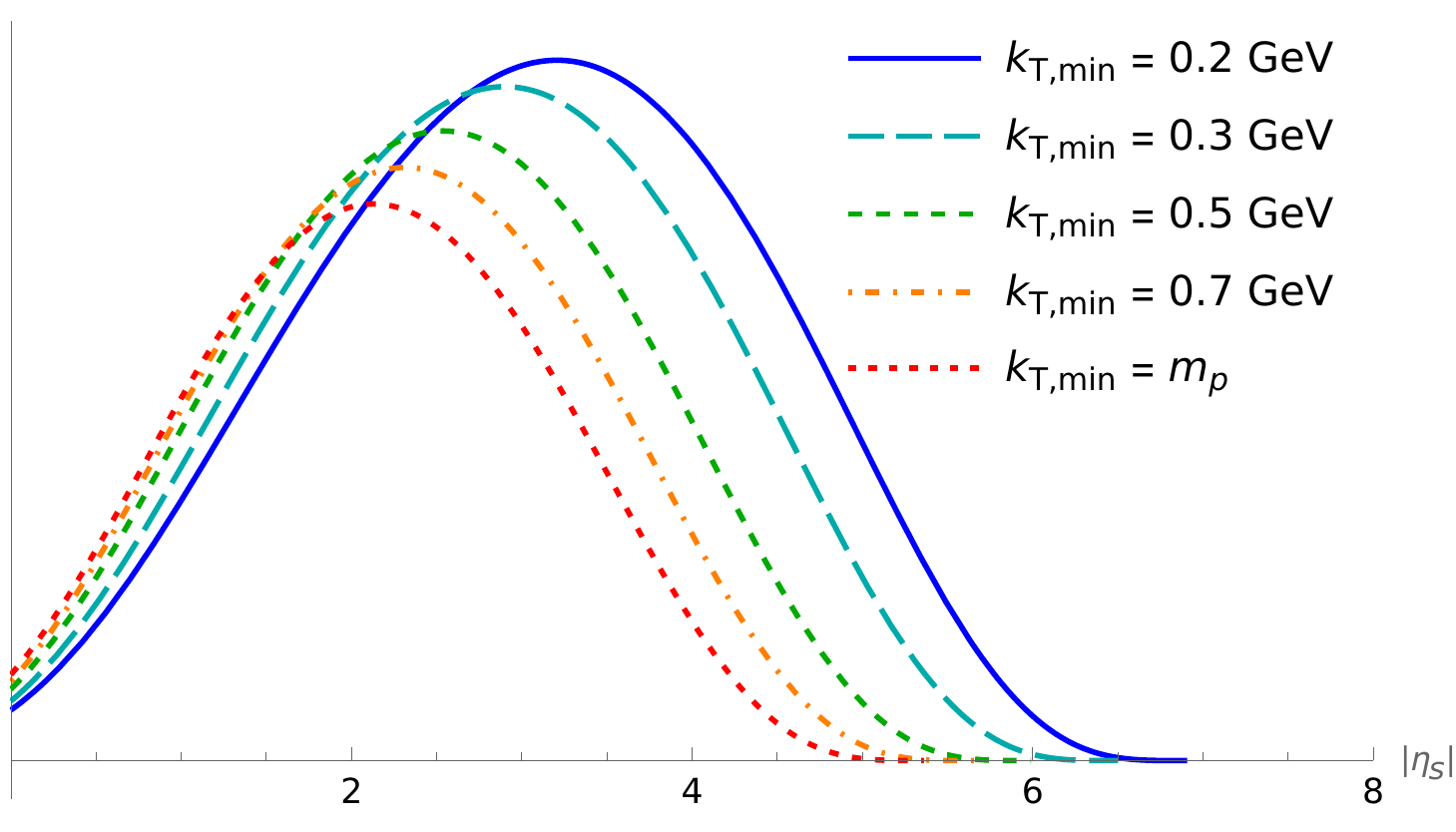}
  \caption
  {
    \label{fig:fragparams}
    Fragmentation profile (Eq.~\eqref{eq:ffrag}) at $\sqrts = \qty{200}{GeV}$ for various values of $\alpha$ (left), $\beta$ (middle), or $\kTmin$ (right), with the other two parameters fixed to arbitrary values.
  }
\end{figure*}

With the model now presented in full, we describe each parameter in detail below and summarize them for reference in Table~\ref{tab:params}:

\begin{itemize}
\item{\textit{Constituent number} ($n_c$).  The number of subnucleonic degrees of freedom in each nucleon.  For $n_c > 1$, each nucleon's thickness is structured as $n_c$ smaller, Gaussian ``hotspots'' located in the vicinity of the nucleon.  If $n_c = 1$, there is no nucleonic substructure: the nucleon's thickness is simply a single Gaussian.}
\item{\textit{Nucleon width} ($w$).  The width in fm of the 2D-Gaussian representation of each nucleon.  With nucleonic substructure, this width influences the localization of the constituents, in that they are placed according to a normal distribution of width $\sqrt{ \frac{w^2 - v^2}{1 - 1 / n_c} }$.  Without substructure, $w$ instead parametrizes the 2D Gaussian shape of the nucleon's thickness. }
\item{\textit{Constituent width} ($v$).  The 2D-Gaussian width in fm of each constituent's thickness contribution, when nucleonic substructure is used.  Note that $v$ must be less than $w$.}
\item{\textit{Form width} ($u$).  The 2D-Gaussian nucleon width in fm for the purpose of computing the probability of collision, distinct from $w$.  The probability is
  \begin{equation}
    P(\bNN)= 1 - \exp \left\lbrace -\sigmainel \left[ \frac{1}{4 \pi u^2} e^{-\bNN^2 / 4 u^2} \right] \right\rbrace\text{,}
  \end{equation}
  where the quantity in square brackets is the overlap of two normalized Gaussians, each of width $u$, when their centers are separated by a distance $\bNN$.  $\sigmainel$ is the inelastic nucleon--nucleon cross section, computed via the expression
  \begin{eqnarray}
    \sigmainel / \mathrm{fm}^2 &=& 3.12 + 0.0735 \ln(\sqrts/{\rm GeV}) \nonumber \\
    && {} - 0.193 [\ln(\sqrts/{\rm GeV})]^2 \label{eq:sigmainel}
  \end{eqnarray}
  which has been fit to Particle Data Group data~\cite[Ch.~53]{ParticleDataGroup:2022pth}.  (In the original \trento{}, $\sigmainel$~was a required parameter since $\sqrts$ was not specified.)
}
\item{\textit{Fluctuation} ($k$).  Determines the shape of the probability distribution~\eqref{eq:betadistro} from which each constituent's fluctuation factor $\gamma_c$ is sampled, with larger $k$ typically resulting in more extreme values (Fig.~\ref{fig:fluctuation}).  Regardless of $k$, the mean of the distribution always remains $\avgxloss$ by construction.}
\item{\textit{Fireball normalization} ($\normfb$).  Specifies the energy scale of the central fireball per nucleon pair.  The energy content of the fireball profile~\eqref{eq:ffb} (centered about $\etascm = 0$) is
  \begin{equation}
    \normfb \int_{-\etasmax}^{+\etasmax} \cosh(\etas) ~ d\etas ~ \ffb(\etas)\text{.}
  \end{equation}
  By energy conservation this integral must be less than $\sqrts$, as the difference will be deposited in the fragmentation regions.  This constraint permits an inequality relating $\normfb$ to $\sqrts$, $\flatness$, $\nu$, and $\kTmin$, but even to the extent it can be expressed in closed form---for instance, when $\flatness = 1$---it is not succinct and so we do not include it here.
}
\item{\textit{Flatness} ($\flatness$).  Allows the central fireball profile to deviate from a Gaussian, becoming more plateau-like with higher values of $\flatness$ (Fig.~\ref{fig:flatness}).  Conceptually, $\flatness$ interpolates between a Gaussian at $f = 1$ and the rectangular function in the $f \to \infty$ limit.}
\item{\textit{Fragmentation shape parameters} ($\alpha$ and $\beta$).  Control the shape of the fragmentation profile, with $\beta + 1$ the power of momentum fraction $x$ and $\alpha$ the exponent of $\ln\left(1/x\right)$ in Eq.~\eqref{eq:ffrag}. The effect of $\alpha$ ($\beta$) on the fragmentation profile is illustrated in the left (middle) panel of Fig.~\ref{fig:fragparams}.}
\item{\textit{Transverse momentum scale} ($\kTmin$).  Specifies the typical transverse momentum of the massless partons composing the initial conditions, which influences the shapes and extents in rapidity of both the central fireball and the fragmentation profiles.The simultaneous effects on the latter are evident in the right panel of Fig.~\ref{fig:fragparams}: with smaller $\kTmin$, the fragmentation profile both extends farther in rapidity due to increasing $\etasmax$ (see Eq.~\eqref{eq:etasmax}), and becomes larger as the exponential decay term of Eq.~\eqref{eq:ffrag} loses strength.}
\end{itemize}

\section{A Simplified Evolution Model}

For the purposes of verification and validation of \trentoDDD{}, we introduce a simplified evolution model that is capable of describing basic observables of ultrarelativistic collisions---such as the rapidity dependence of the charged-particle multiplicity---that do not depend strongly on the details of a hydrodynamic evolution.  This will allow us to connect the \trentoDDD{} initial state to a set of final-state observables for comparison.  In particular, our simplified evolution model accounts for the expansion that manifests itself as longitudinal smoothing and broadening of the charged-particle (pseudo-)rapidity distribution relative to the initial-state energy distribution.

We neglect any pre-equilibrium phase and assume that the evolution of the system is well described by hydrodynamics. For the sake of simplicity and computational expediency we limit ourselves to (1+1)D linearized ideal hydrodynamics. Our evolution model has thus the following components:
\begin{itemize}
\item{\textit{(1+1)D linearized hydrodynamics.}  The Bjorken-frame energy density profile $d\ET / \tau d\etas$ is evolved by inviscid relativistic hydrodynamics, in the longitudinal and proper-time dimensions only, using an ideal-gas equation of state and a first-order approximation in velocity.}
\item{\textit{Transverse Gaussianization.}  After hydrodynamic evolution, the time-evolved longitudinal energy density is assumed to have a transversely Gaussian distribution for the purpose of constructing a freeze-out hypersurface in (3+1)D and performing Cooper--Frye particlization~\cite{Cooper:1974mv}.}
\item{\textit{No afterburner.}  The particle spectra generated via Cooper--Frye particlization are treated directly as observables; hadronic rescattering and decays are not simulated.  In order to roughly account for their effects on the observable, we multiply our charged hadron yields by a factor of two and assume that the model parameters will provide further compensation when we compare to physical data in Sec.~\ref{sec:results}.}
\end{itemize}
We shall describe the details of this simplified evolution model in the following sections.

It is worth emphasizing that because the \emph{simplified hydrodynamics model} has no transverse dynamics, the present work only considers observables such as particle multiplicity and not flow anisotropies or the like.  This is not a limitation of \trentoDDD{} itself; the \emph{initial conditions} express highly intricate structure (as illustrated in Fig.~\ref{fig:IC3D}) capable of driving full (3+1)D dynamics, but for this first study, we choose to ignore most of that structure in exchange for a computationally cheap (1+1)D hydrodynamic evolution.

\subsection{\protect\trentoDDDalt{} Output}
\label{sec:output}

Schematically, the core of a heavy-ion collision simulation is hydrodynamic evolution, the inputs of which are the unique components of the energy--momentum tensor $\Tmunu(\tau_0, \xTvec, \etas)$ (as well as an equation of state and various parameters such as $\tau_0$).  In contrast, the output of \trentoDDD{} is the Bjorken-frame energy distribution $\tauIC ~ \epsIC(\xTvec, \etas; \tauIC)$, with units $\GeVperfmsq$ (one must choose a $\tauIC$ to divide by, as mentioned in Sec.~\ref{sec:3Dansatz}, to get a density in $\GeVperfmcu$), discretized onto a 3D grid with coordinates $(x, y, \etas)$.  For later reference, we note that the grid is longitudinally divided into a specified number of \term{rapidity slices}---$x$--$y$ grids of cells having a common $\etas$ coordinate---that span the interval $[-\etasgridmax, +\etasgridmax]$, with
\begin{equation}
\etasgridmax = \arccosh\left(\frac{\sqrts / 2}{\LambdaQCD}\right)\text{,}
\end{equation}
where $\LambdaQCD = \qty{0.2}{GeV}$.

Thus \trentoDDD{} supplies the energy contribution to $\Tmunu$, but it has nothing to say regarding pressure or flow beyond what can be derived from the energy.  This situation is the same as with \trento{}~\cite[Ch.~5.1.6]{Moreland:2019szz}, where a common solution has been to interpose a free-streaming pre-equilibrium stage~\cite{Broniowski:2008qk,Liu:2015nwa}, although doing so is understood to be somewhat unphysical, and alternative approaches have been studied~\cite{Keegan:2016cpi,Kurkela:2018vqr,NunesdaSilva:2020bfs}.  In future work we will connect \trentoDDD{} to hydrodynamics via (3+1)D pre-equilibrium dynamics; for our present goals of introducing and validating the model, and demonstrating its flexibility, we instead choose to simply omit the pre-equilibrium stage.  Consequently there is no local flow---i.e., $u^\mu(\tau_0) = (1, \vec{0})$---at the hydrodynamization time $\tau_0 = \tauIC$.

\subsection{Linearized Hydrodynamics}
\label{sec:LH}

As mentioned earlier, it is not expected that \trentoDDD{} initial conditions would compare well to final-state observables across a broad range in rapidity without longitudinal evolution.  The present simplified hydrodynamics model, dubbed LH, was devised to approximate such an evolution at small computational cost.  Full details of its derivation can be found in Appendix~\ref{app:LH}.

The practical purpose of the LH model is to compute the energy and longitudinal momentum profiles $\LHe(\tau, \etas)$ and $\LHpz(\tau, \etas)$ as functions of proper time.  Here the term ``profile'' denotes a longitudinal density (i.e., with respect to $\etas$), as LH operates in (1+1)D---perhaps the most consequential simplification we impose.  The initial conditions of these quantities are
\begin{align}
\LHe(\tau_0, \etas) & = \frac{d\ET}{\tau_0 ~ d\etas}(\tau_0, \etas) \\
& = \int d^2\xTvec ~ \epsIC(\xTvec, \etas; \tau_0)\text{,} \\
\LHpz(\tau_0, \etas) & = 0\text{,}
\end{align}
where $\epsIC$ represents the 3D initial conditions (Eq.~\eqref{eq:epsICdET}) and $\tau_0$ is a hydrodynamization time that must be specified.  (In the analysis of Sec.~\ref{sec:analysis}, we treat $\tau_0$ as a free parameter.)

The hydrodynamic formulation is made simpler still by neglecting viscosity, so that $\Tmunu(\tau, \etas)$ takes the form
\begin{equation}
\Tmunu = (\LHe + P) u^\mu u^\nu - P g^{\mu\nu}\text{,}
\end{equation}
with $u^\mu(\tau, \etas)$ the (two-component) 4-velocity representation of longitudinal flow $\LHvz(\tau, \etas)$, defined
\begin{equation}
u^\mu = \frac{1}{\sqrt{1 - \LHvz^2}} \left(1, \frac{\LHvz}{\tau}\right)\text{,}
\end{equation}
and $g^{\mu\nu}$ the metric (Eq.~\eqref{eq:LHg}, which has signature $(+{}-)$).  To eliminate the pressure $P$ in favor of the energy density $\LHe$, we simply assume the ideal-gas equation of state,
\begin{equation}
P = \cssq ~ \LHe\text{.}
\end{equation}
($\cssq$ is the squared speed of sound and will take the value $1/3$.)  Finally, we expand $\Tmunu$ only to first order in $\LHvz$, which gives
\begin{equation}
\Tmunu = \begin{pmatrix}
1 & \left(1 + \cssq\right) \frac{\LHvz}{\tau} \\
\left(1 + \cssq\right) \frac{\LHvz}{\tau} & \frac{c_s^2}{\tau^2}
\end{pmatrix} \LHe\text{.} \label{eq:LHTmunu}
\end{equation}
Applied to the above, the continuity equation Eq.~\eqref{eq:LHcont} then admits a solution that can be evaluated at arbitrary $\tau$.

In addition to being fast to evaluate, LH allows an efficient determination of the \term{freeze-out time} $\tauFO$, after which the entire system has cooled to below the \term{switching temperature} $\TFO$, to arbitrary precision through a ``binary search''-like approach.  However, before the usual freeze-out paradigm can be applied, we must reconstitute our (1+1)D system as a (3+1)D energy distribution.  We present a crude solution below.

\subsection{Transverse Gaussianization}
\label{sec:TG}

In employing (1+1)D hydrodynamics we chose to eliminate the transverse dimensions for simplicity, but to perform the usual Cooper--Frye particlization procedure and obtain particle yields, we need to return to (3+1)D.  The reason is that Cooper--Frye requires a freeze-out hypersurface, a (2+1)D isotherm which demands a switching temperature $\TFO$---or an equivalent 3D energy density, $\epsFO$---and therefore an extent in 3D space.

To that end, we assume the energy $\LHe(\tau, \etas) ~ \tau ~ d\etas$ at any given $\etas$ is transversely distributed as a 2D Gaussian, such that
\begin{equation}
\epsG(\tau, r, \phi, \etas) = \LHe(\tau, \etas) \frac{1}{2 \pi \sigmaG^2} e^{-r^2 / 2 \sigmaG^2}\text{,}
\end{equation}
with $r$ the (radial) distance from the $z$~axis, $\phi$ the azimuthal angle, and $\sigmaG$ a Gaussian width.  (We compute $\sigmaG$ for each event as the energy-weighted average of radii enclosing $1 - e^{-1/2} \approx 39.3\%$ of the energy in each rapidity slice of the initial conditions.  For simplicity it is assumed to not vary with $\tau$ or $\etas$.)  The Gaussian is always centered on the $z$~axis, irrespective of how the transverse center of mass actually varies among rapidity slices in the initial conditions.

The freeze-out hypersurface is defined by $\epsG(\tau, r, \phi, \etas) = \epsFO$.  The solution is azimuthally symmetric, so we can express the hypersurface in terms of the \term{freeze-out radius}
\begin{equation}
\rFO(\tau, \etas) = \sqrt{ 2 \sigmaG^2 \ln \left\lbrace \frac{\LHe(\tau, \etas)}{2 \pi \sigmaG^2 \epsFO} \right\rbrace }\text{.}
\end{equation}
(For compactness, we have omitted some clipping that ensures $\rFO$ goes to zero rather than becoming complex when a rapidity slice is everywhere below $\epsFO$.)  Incidentally, we observe that this Gaussian construction inevitably manifests a ``corona''~\cite{Werner:2007bf} of subthreshold initial energy beyond $\rFO(\tau_0, \etas)$ which may be presumed lost.

In the present study, we discretize the freeze-out hypersurface $\rFO(\tau, \etas)$ into $\netas$ rapidity slices of uniform width
\begin{equation}
\Delta\etas = \frac{2 ~ \etasgridmax}{n_{\etas}}
\end{equation}
and $\ntau$ timesteps of variable duration
\begin{equation}
\Delta\tau(\tau) = \frac{1}{\ntau} \left(\frac{\tauFO}{\tau_0}\right)^\frac{\tau - \tau_0}{\tauFO - \tau_0} \ln \left(\frac{\tauFO}{\tau_0}\right)\text{,}
\end{equation}
where $\tauFO$ is the freeze-out time (determined in advance; see Sec.~\ref{sec:LH}), the earliest time at which $\rFO(\tauFO, \etas) = 0$ is true for all $\etas$.  The timestep durations are uniform on a \emph{logarithmic scale}, being shortest at early times.  Heuristically this allows the hydrodynamic evolution, which generally takes the form of a decay steepest at early times, to be better captured given a fixed number of timesteps.

With the freeze-out hypersurface $\rFO$ determined for all $\tau$ and $\etas$, we can now compute particle production.

\subsection{Cooper--Frye}
\label{sec:CF}

The Cooper--Frye particlization procedure~\cite{Cooper:1974mv} computes the expected particle emission from a thermal hypersurface.  Cast in terms of pseudorapidity $\eta = \arctanh\left(p_z / p\right)$ to best match available experimental data, the Cooper--Frye integral reads
\begin{align}
\frac{dN_i}{d\eta}(\eta) = \frac{1}{\hbar^3} \int_0^\infty d\pT \int_0^{2\pi} \pT ~ d\phi_p ~ \times \nonumber \\
\sqrt{1 + \frac{m_i^2}{\pT^2 \cosh^2 \eta}} ~ \times \nonumber \\
\int_\Sigma d^3\Sigma_\mu p^\mu ~ f(x, p)\text{.}
\end{align}
On the left, $dN_i/d\eta$ represents the pseudorapidity number density at a given $\eta$ of particles of species~$i$, each having mass~$m_i$.  The first two integrals on the right are over transverse momentum in cylindrical coordinates, while the third integrates over elements $d^3\Sigma$ of the freeze-out hypersurface, with $d^3\Sigma_\mu$ being the surface normal 4-covector and $x$ the position 4-vector of the surface element.  $p$ is the 4-momentum
\begin{equation}
p^\mu = \begin{pmatrix}
\sqrt{m_i^2 + \pT^2 + \pT^2 \sinh^2 \eta} \\
\pT \cos \phi_p \\
\pT \sin \phi_p \\
\pT \sinh \eta
\end{pmatrix}\text{,}
\end{equation}
with $\pT \sinh\left(\eta\right)$ expressing longitudinal momentum.  The factor $\hbar^{-3} \approx \left(0.19733~\mathrm{GeV}\cdot\mathrm{fm/c}\right)^{-3}$ remains from the Cooper--Frye equation's original form as a momentum distribution.  Lastly, $f(x, p)$ is the single-particle distribution function
\begin{equation}
f(x, p) = \frac{g_i}{(2 \pi)^3 \left( \exp\left\lbrace u'_\mu p^\mu / \TFO \right\rbrace \pm 1 \right)},
\end{equation}
where $g_i$ is a degeneracy factor for particles of species~$i$, $(\pm)$ is $(+)$ for fermions and $(-)$ for bosons, $\TFO$ is the switching temperature (Secs.~\ref{sec:LH}, \ref{sec:TG}), and $u'$ is the \emph{lab-frame} hydrodynamic flow 4-velocity, obtained through boosting the local flow velocity $v_\mathrm{LH}$ (Sec.~\ref{sec:LH}) by $\tanh\left(-\etas\right)$, as follows:\footnote{
  This prescription maintains our earlier identification of the lab frame as the nucleon--nucleon center of mass frame; however, even in colliding-beam experiments this is not always the case.  See for example Ref.~\cite{ATLAS:2015hkr}.
}
\begin{align}
v'_\mathrm{T} & = \frac{1 - \tanh^2 \etas}{1 + \LHvz \tanh \etas} \vT \\
v'_z & = \frac{\LHvz + \tanh \etas}{1 + \LHvz \tanh \etas} \LHvz \\
u' & = \frac{1}{\sqrt{1 - (v'_\mathrm{T})^2 - (v'_z)^2}} \begin{pmatrix}
1 \\
v'_\mathrm{T} \cos \phi_u \\
v'_\mathrm{T} \sin \phi_u \\
v'_z
\end{pmatrix}\text{.}
\end{align}
Here, $\phi_u$ represents the azimuthal angle of the flow, but we have only included it for the sake of generality; we will exploit cylindrical symmetry\footnote{
  For alternative cylindrically symmetric treatments, see Chaudhuri~\cite[Ch.~7.3.1]{Chaudhuri:2012yt} and Schenke \textit{et al.}~\cite{Schenke:2010nt}.  A primary difference is that we have expressed the hypersurface in terms of the freeze-out radius as a function of time, rather than the freeze-out time as a function of radius---i.e., $\rFO(\tauFO, \etas)$ versus $\tau_f(r_\perp)$ or $\tau_f(x, y, \etas)$.
}
to eliminate the azimuthal components of both $d^3\Sigma_\mu$ and $u^\mu$ from their respective dot products.

The transverse flow velocity $\vT$ is of course not provided by (1+1)D hydrodynamics; rather, we use the approximate expression for the cylindrically symmetric case obtained in Ref.~\cite{Paquet:2022wgu}:\footnote{
  Note that Ref.~\cite{Paquet:2022wgu} specifies the width $\sigma_0$ of the \emph{temperature} transverse Gaussian.  Since we are working with energy density, and $\varepsilon \propto T^4$ by our choice of equation of state, we have replaced $\sigma_0$ with $2\sigmaG$ in constructing Eq.~\eqref{eq:uT}.
}
\begin{equation}
\uT = \gamma \vT = \frac{r_f}{4\sigmaG^2} \frac{ \tau - \tau_0 \left(\frac{\tau}{\tau_0}\right)^{\cssq} }{ \left(1 - \cssq\right) \left( 1 + \frac{\tau^2}{8\sigmaG^2} \right) }\text{.} \label{eq:uT}
\end{equation}
Further derivation of our Cooper--Frye implementation is provided in Appendix~\ref{app:CF}.  Most of the details are not needed for the present discussion, although we should briefly mention the derivatives that arise in the definition of the surface normal (Eq.~\eqref{eq:CFdSigma}).  Specifically, $d^3\Sigma_\mu$ contains factors of $\partial_\tau \rFO$ and $\pdetas \rFO$ that will need to be computed from the discretized hypersurface.  One simple option is finite differences or stencils; another is an interpolator that accommodates differentiation, such as a 2D (meaning $\tau$--$\etas$) cubic spline, although one should be mindful of the Gibbs phenomenon when using higher-order interpolation.

With these ingredients, all that remains is to perform a numerical integration over momenta for each hypersurface element, and to sum over the hypersurface elements, for each particle species.  The result is the expected (average) pseudorapidity-dependent spectrum for the particle.

\section{Model Validation}
\label{sec:analysis}

In this section we shall demonstrate \trentoDDD{}'s viability as an initial-conditions model through a series of tests that rely on our simplified evolution model in order to account for the broadening of the final-state charged-particle rapidity distributions due to the expansion of the QGP. Utilizing this evolution model we perform the following tests:
\begin{enumerate}
\item A closure test to demonstrate the validity of our analysis framework.
\item A large-scale calibration of our simplified evolution model, utilizing rapidity distributions of charged hadrons generated by a high-fidelity (3+1)D relativistic viscous hydrodynamics model with \trentoDDD{} initial conditions as mock data. This calibration aims to demonstrate the ability of our simplified evolution model to act as a stand-in for a high-fidelity model and recover salient features of a (known) \trentoDDD{} initial state.
\item A calibration of our \trentoDDD{} initial stage plus simplified evolution model to actual experimental data, to demonstrate \trentoDDD{}'s ability to provide intial conditions that can result in realistic final-state distributions.
\end{enumerate}
The following sections describe our methodology and present the results of each test in succession.

\subsection{Bayesian Calibration}

The context in which a parametric model such as \trentoDDD{} often appears is that of the ``inverse problem,'' which schematically might be written
\begin{equation}
\mathrm{Parameters} = \mathrm{Model}^{-1}\left( \mathrm{Data} \right)\text{,}
\end{equation}
in contrast to the ``forward problem'' $\mathrm{Data} = \mathrm{Model}\left( \mathrm{Parameters} \right)$.  In other words, we typically wish to determine which parameter values cause simulated data to best match experimental data, evaluate how well they agree, and then extract physical insight from those parameter values or use them to make novel predictions, or simply investigate the performance of the model.

Of course, reversing the models under consideration, from observables back to parameter values, is not directly possible.  One might be tempted to consider an iterative approach in which the model is repeatedly evaluated with different parameter values and the results compared to expectations, but two factors conspire to render this infeasible as such: the computational cost of generating a statistically sufficient number of events, and the high dimensionality of the parameter space.

Fortunately both challenges have well-established solutions.  To address the computational cost, we train a Gaussian process (GP) to ``emulate'' the model as in Refs.~\cite{Petersen:2010zt,Novak:2013bqa} (and many subsequent works), essentially interpolating among and extrapolating from the \term{training points} in parameter space at which we have run the model, to guess what it would predict at points where we have not.  The set of training points is termed the \term{design}; we use a common choice of layout called a ``Latin hypercube,'' which offers far superior performance for a given computational budget versus a regular grid.  While it is also common to apply dimension reduction, especially principal component analysis (PCA), to the observables, we obtain better results when using the multiple-output GP implemented in \texttt{Scikit-learn}~\cite{Pedregosa:2011ork} without PCA.  In any case, the GP (or in other works, the combination of PCA and multiple GPs) is more generically referred to as an \term{emulator}, and its sole purpose is to inexpensively predict the observables that the actual model would produce at a given parameter point, while qualifying that prediction with an \term{emulator uncertainty}.

As for the challenge of dimensionality, the Markov Chain Monte Carlo algorithm offers a highly efficient solution for exploring the parameter space.  Our goal is to estimate the \term{posterior} function, which can be coarsely described as quantifying how well the simulated observables match experiment, and moreover we wish to locate its maximum.  The posterior is a function of the model parameters, and we have already established that there are too many parameters to fully scan over the space, so approximating the posterior with a complete, $N$-dimensional histogram is out of the question.  What MCMC instead provides is a ``point cloud''---termed the \term{chain} due to how it is constructed---in which points concentrate around the maxima or ideally the maximum.  Adequately representing the posterior in this way might require millions of evaluations, and since we have also established that the model is too costly to run so many times, successfully using MCMC depends on having a fast emulator.  By combining emulation and MCMC,\footnote{
  Perhaps the earliest realization of this approach in heavy-ion physics is Ref.~\cite{Novak:2013bqa}, although it has since become standard practice.
} we solve both problems and can proceed with the analysis.

The last few paragraphs only sketch out the general concepts and terminology of our analysis approach---a detailed discussion is beyond the scope of this article.  For excellent introductions to the Bayesian analysis toolkit commonly used in heavy-ion physics, we refer the reader to Refs.~\cite{Bernhard:2018hnz,Moreland:2019szz}.

\subsection{Analysis Methodology}

\begin{table*}
  \centering
  \begin{tabular}{|c|c|c|l|}
    \hline
    \textbf{System} & \textbf{$\sqrts$} & \textbf{Data} & \textbf{Centrality Bins} \\
    \hline
    Pb--Pb & $\qty{5.02}{TeV}$ & ALICE~\cite{ALICE:2016fbt} & 0--5\%, 5--10\%, 10--20\%, 20--30\%, 30--40\%, 40--50\%, 50--60\%, 60--70\%, 70--80\% \\
    \hline
    $p$--Pb & $\qty{5.02}{TeV}$ & ATLAS~\cite{ATLAS:2015hkr} & 0--1\% 1--5\%, 5--10\%, 10--20\%, 20--30\%, 30--40\%, 40--60\%, 60--90\% \\
    \hline
    Au--Au & $\qty{200}{GeV}$ & PHOBOS~\cite{Back:2002wb} & 0--6\%, 6--15\%, 15--25\%, 25--35\%, 35--45\%, 45--55\% \\
    \hline
    $d$--Au & $\qty{200}{GeV}$ & PHOBOS~\cite{PHOBOS:2004fzb} & 0--20\%, 20--40\%, 40--60\%, 60--80\% \\
    \hline
  \end{tabular}
  \caption
  {
    \label{tab:systems}
    Collision systems and energies selected in this work.  The model-to-data comparison (Sec.~\ref{sec:results}) seeks to match the referenced experimental data; the closure test (Sec.~\ref{sec:closure}) and cross-model comparison (Sec.~\ref{sec:modeltomodel}) do not, but they nonetheless use similar specifications, such as the experimentally-defined centrality bins.
  }
\end{table*}

In this section we specify in detail the steps undertaken in the analyses that follow.  We target the four collision systems summarized in Table~\ref{tab:systems}, attempting to discover or recover a single set of parameter values that simultaneously reproduces all four datasets, although two quantities, the fireball normalization $\normfb$ and the hydrodynamization time $\tau_0$, are split into multiple parameters according to energy and collision system respectively.  We focus on a single observable, the charged hadron ($\pi^\pm, K^\pm, p, \bar{p}$) spectrum versus pseudorapidity, denoted $d\Nch/d\eta$.  This choice reflects limitations imposed \emph{by our simplified hydrodynamic and post-hydrodynamic stages}; we stress that there is no such limitation inherent in \trentoDDD{}.

\begin{table}
  \centering
  \begin{tabular}{|p{0.55 \columnwidth}|c|c|}
    \hline
    \textbf{Parameter} & \textbf{Symbol} & \textbf{Range} \\
    \hline
    Form width [fm] & $u$ & 0.35 -- 1.0 \\
    \hline
    Nucleon width [fm] & $w$ & 0.35 -- 1.0 \\
    \hline
    Constituent number & $n_c$ & 2.0 -- 20.0 \\
    \hline
    Structure & $\chi$ & 0.2 -- 0.9 \\
    \hline
    Transverse mom. scale [GeV] & $\kTmin$ & 0.2 -- 0.9 \\
    \hline
    Shape parameter & $\alpha$ & 3.0 -- 5.0 \\
    \hline
    Shape parameter & $\beta$ & $-0.5$ -- 1.5 \\
    \hline
    Fireball norm. [GeV] \newline ($\sqrts = \qty{200}{GeV}$) & $N_{200}$ & 1.0 -- 15.0 \\
    \hline
    Fireball norm. [GeV] \newline ($\sqrts = \qty{5.02}{TeV}$) & $N_{5020}$ & 15.0 -- 30.0 \\
    \hline
    Fluctuation & $k$ & 0.1 -- 0.6 \\
    \hline
    Flatness & $\flatness$ & 1.0 -- 2.5 \\
    \hline
    Hydrodyn. time [fm/c] \newline (Pb--Pb $\qty{5.02}{TeV}$) & $\tau_{0,\mathrm{Pb}}$ & 0.1 -- 1.5 \\
    \hline
    Hydrodyn. time [fm/c] \newline ($p$--Pb $\qty{5.02}{TeV}$) & $\tau_{0,p}$ & 0.1 -- 1.5 \\
    \hline
    Hydrodyn. time [fm/c] \newline (Au--Au $\qty{200}{GeV}$) & $\tau_{0,\mathrm{Au}}$ & 0.1 -- 1.5 \\
    \hline
    Hydrodyn. time [fm/c] \newline ($d$--Au $\qty{200}{GeV}$) & $\tau_{0,d}$ & 0.1 -- 1.5 \\
    \hline
    Overall scale & $\Nscale$ & 0.8 -- 2.0 \\
    \hline
  \end{tabular}
  \caption{
    \label{tab:analysisparams}
    The parameters used in the analyses, listed in the order in which they appear in the parameter posterior plots.  Note that the parametrization is somewhat different from that represented in Table~\ref{tab:params}, for reasons explained in the text.  The rightmost column lists the prior for each parameter.
  }
\end{table}

The 16 model parameters we intend to estimate appear in Table~\ref{tab:analysisparams}.  There are a couple reparametrizations with respect to \trentoDDD{}, and a a couple new parameters pertaining to the simplified evolution model:

\begin{itemize}
\item{\textit{Structure} ($\chi$) \textit{versus constituent width} ($v$).  The constituent width is mildly inconvenient as a free parameter since \trentoDDD{} requires $v < w$.  Therefore, instead of directly including $v$, we follow Ref.~\cite{Nijs:2020ors} in devising a ``structure'' parameter $\chi$ that expresses the constituent width in terms of the nucleon width.  Doing so allows us to maintain a simpler, ``hyper-rectangular'' parameter space, avoiding a triangle of invalidity in the $w$--$v$ plane that would result for all $v \geq w$.  We define the structure parameter
  \begin{equation}
  \chi \equiv \frac{v}{w ~ n_c^{1/4}}
  \end{equation}
such that it also depends on the number $n_c$ of constituents.}
\item{\textit{Energy-dependent fireball normalizations} ($N_{200}, N_{5020}$).  The current implementation of \trentoDDD{} computes the fireball normalization $\normfb$ as a function of $\sqrts$ with parameters $N_\mathrm{mid}$ and $p_\mathrm{mid}$:
  \begin{equation}
  \normfb(\sqrts) = N_\mathrm{mid} ~ m_p ~ \left( \frac{\sqrts}{m_p} \right)^{p_\mathrm{mid}}
  \end{equation}
(where $m_p$ is the nucleon mass).  However, since we are only considering two energies, we find it expedient to make $\normfb$ directly a free parameter at each energy---i.e., $N_{200}$ and $N_{5020}$ in lieu of $N_\mathrm{mid}$ and $p_\mathrm{mid}$.}
\item{\textit{Hydrodynamization time} ($\tau_0$).  A common though coarse assumption in modeling ultrarelativistic heavy-ion collisions is that the collision system abruptly enters the hydrodynamic regime at a certain proper time $\tau_0$.  This time constitutes an additional free parameter in the analysis, although it would be too unphysical to further assume that the time is the same regardless of collision energy and system size, so we instead maintain a separate $\tau_0$ per system.}
\item{\textit{Overall scale} ($\Nscale$).  The initial conditions generated by \trentoDDD{} constitute an energy distribution and ensure that the total participant energy $\Npart \sqrts / 2$ is conserved.  Nonetheless, we have included an overall normalization parameter by which the initial conditions are multiplied to compensate for shortcomings in the simplified hydrodynamics and particlization models.  Ideally a value of unity will be strongly favored.}
\end{itemize}

There are other quantities that are nominally parameters but that we have fixed in the present analyses:

\begin{itemize}
\item{\textit{Narrowing term} ($\nu$).  The rough form of the central fireball profile~\eqref{eq:ffb} is that of a Gaussian of width $\sqrt{\etasmax - \nu}$.  To simplify the analysis, we fix $\nu = 3$ in the present work.}
\item{\textit{Tapering exponents}.  The tapering factor in Eq.~\eqref{eq:ffb} is quartic, but it need not be so: with some restrictions, the inner and outer exponents could be considered parameters.  Although the tapering factor's effect on the fireball profile is small (Fig.~\ref{fig:flatness}), its effect on the total energy density---which goes as $\cosh(\etas)$---can be significant, and so too its effect on the interpretation of $\normfb$.  Nevertheless, we leave such considerations for future work.}
\item{\textit{Minimum nucleon--nucleon separation} ($d_\mathrm{min}$).  \trento{} allows for a minimum distance to be maintained among nucleons, analogous to a hard-core repulsion.  However, the parameter naturally has no relevance for $H^+$, it is disregarded for the deuteron, and it tends to be poorly constrained for large nuclei, so we do not use it in this work.}
\item{\textit{Switching temperature} ($\TFO$).  The switching temperature is an important and physically relevant parameter in the study of quark--gluon plasma, but given the simplified evolution and particlization models and our limited selection of observables, we do not expect the analysis to be sufficiently sensitive to it.  Therefore we simply set $\TFO = \qty{150}{MeV}$.}
\item{\textit{Switching energy density} ($\epsFO$).  Although largely fixed by our choice of $\TFO$, $\epsFO$ has limited independence insofar as we are free to choose an equation of state (notwithstanding our assumption of the ideal-gas equation of state in Sec.~\ref{sec:LH}).  Using the HotQCD equation of state~\cite{HotQCD:2014kol}, we obtain for the above choice of $\TFO$ a corresponding $\epsFO = 0.234~\GeVperfmcu$.}
\end{itemize}

Through initial testing, we arrived at reasonable ranges of values to allow the parameters to take, as listed in Table~\ref{tab:analysisparams}.  In the language of Bayesian analysis, these ranges constitute the \term{prior}, a distribution that captures our initial beliefs about the parameters and is used in computing the posterior.  A common and simple choice is the \textit{uniform prior}, often in the form of a range, outside of which the distribution is zero, and inside of which it is positive and constant.  The outer product of these ranges defines a ``hyper-rectangular'' region of parameter space to which MCMC is confined, and outside of which the posterior vanishes.

With the parameters and ranges decided, the next step is to generate a design for training the emulator.  We created a maximin Latin hypercube~\cite{Ba:2015} comprising 1,000 design points, eight of which were excluded because their specific combinations of parameter values would have caused the central fireball to exceed the total collision energy (i.e., $\avgxloss > 1$).  At each of the remaining 992 design points, we generated 5,000 \trentoDDD{} events per collision system using that point's associated initial-state parameter values.  In the case of $p$--Pb collisions at $\sqrts = \qty{5.02}{TeV}$, we shifted the midrapidity of the averaged events into the $p$-going direction by 0.465 units of $\etas$, to compensate for an asymmetry in the proton and lead beams' energies at the LHC~\cite{ATLAS:2015hkr}.  The events were then sorted according to the Bjorken-frame energy in a specific range of $\etas$, partitioned into centrality bins (see Table~\ref{tab:systems}), and averaged.  Explicitly, the quantity we used for centrality selection is the integral
\begin{equation}
\int_a^b d\etas ~ \frac{d\ET}{d\etas}(\etas)\text{,}
\end{equation}
with $a$ and $b$ delimiting a spacetime rapidity interval as listed in Table~\ref{tab:etasforcent}.  Finally, the averaged events (one per centrality bin, per collision system, per design point) were supplied to LH,\footnote{
  Hydrodynamically evolving an event-averaged initial condition is known as ``single-shot'' hydrodynamics.  (For an examination of the practice, see e.g. Ref.~\cite{Qiu:2011iv}.)  In the present work, the reduction of longitudinal structure in the initial conditions due to averaging is deemed an acceptable price for the tremendous reduction in overall hydrodynamic computation time.
}
and the evolved profiles were particlized to produce a charged-hadron spectrum per bin, system, and design point.

\begin{table*}
  \centering
  \begin{tabular}{|c|c|c|p{4.5in}|}
    \hline
    \textbf{System} & \textbf{$\sqrts$} & \textbf{$\etas$ Range} & \textbf{Rationale} \\
    \hline
    Pb--Pb & $\qty{5.02}{TeV}$ & $(+2.2, +4.4)$ & Derived from the union of the coverages in $|\eta|$ of the ALICE V0-A and V0-C sub-detectors~\cite{ALICE:2016fbt} (with the sign ignored due to symmetry).  The range is narrowed to increase the relative importance of the interval of overlap. \\
    \hline
    $p$--Pb & $\qty{5.02}{TeV}$ & $(-4.9,-3.1)$ & Based on the $\eta$ coverage of the Pb-going forward calorimeter of ATLAS~\cite{ATLAS:2015hkr}. \\
    \hline
    Au--Au & $\qty{200}{GeV}$ & $(+3.0,+4.5)$ & Based on the $\eta$ coverage of the paddle counters used in PHOBOS to estimate $\langle\Npart\rangle$~\cite{Back:2002wb}. \\
    \hline
    $d$--Au & $\qty{200}{GeV}$ & $(-5.4,-3.0)$ & Based on the $\eta$ coverage of the Au-going ring detector of PHOBOS~\cite{PHOBOS:2004fzb}. \\
    \hline
  \end{tabular}
  \caption
  {
    \label{tab:etasforcent}
    Spacetime rapidity intervals used to determine \trentoDDDalt{} event centrality in the analyses of Sec.~\ref{sec:analysis}, listed for each collision system.
  }
\end{table*}

The pseudorapidities at which the Cooper--Frye formula was evaluated are taken directly from experimental data: 34 values of $\eta \in [-3.5, +5.0]$ for Pb--Pb, 54 values of $\eta \in [-2.7, +2.7]$ for $p$--Pb, and 54 values of $\eta \in [-5.3, +5.3]$ for Au--Au and $d$--Au.  The experiments' centrality selection procedures, on the other hand, are not as easily reproduced.  For one, we are not attempting to simulate all components of the apparatuses; centrality criteria specifying tracks and calorimeter measurements, for instance, have no direct counterparts in the present model.  Moreover, our choice of single-shot hydrodynamics forces us to determine centrality before the hydrodynamic evolution---i.e., according to the initial $d\ET/d\etas$\footnote{
  The sampled impact parameter, though available, would not correlate well with the experimentally-defined centrality in $p$--Pb and $d$--Au collisions, and so we do not consider it in this work.
}
rather than any final-state quantity in terms of pseudorapidity.  Instead, as mentioned above, we settle for integrating $d\ET/d\etas$ over a range of $\etas$ resembling the ranges of $\eta$ chosen by the various experiments (Table~\ref{tab:etasforcent}), and rank the centrality of events accordingly.

The entire process described above is essentially how observables are computed for the design points.  After completing this procedure, we trained one emulator per collision system and then ran MCMC, specifically the {\tt emcee} implementation~\cite{Foreman-Mackey:2012any}, for all four systems simultaneously to generate the posterior.  Our analysis code is an adaptation of the JETSCAPE {\tt STAT} package~\cite{JETSCAPE:STAT}, using a squared-exponential kernel plus a white noise kernel.

\subsection{Closure Test}
\label{sec:closure}

Before attempting a parameter estimation targeting experimental data, a reasonable test is to try to recover parameter values we already know---that is,
\begin{equation}
\mathrm{Model}^{-1}\left( \mathrm{Model}(\mathrm{Parameters}) \right) \stackrel{?}{=} \mathrm{Parameters}\text{.}
\end{equation}
This is called a \term{closure test}.  Specifically, we choose a sensible set of parameter values, generate model data to serve as simulated experimental data, and then perform the usual analysis: generate a design, generate simulated data at the training points, train an emulator, and use MCMC to estimate the posterior, only with simulated instead of real data as the target.  We will be interested in ascertaining both \term{closure}, the ability to recover the chosen parameter values, and \term{constraint}, the extent to which incorrect possibilities are excluded.  To be clear, other than the simulated data targeted, a closure test proceeds exactly as a model-to-data comparison, and so its results will be suggestive of the best-case performance of the model and analysis versus real data.

\begin{figure*}
  \includegraphics[width=\textwidth]{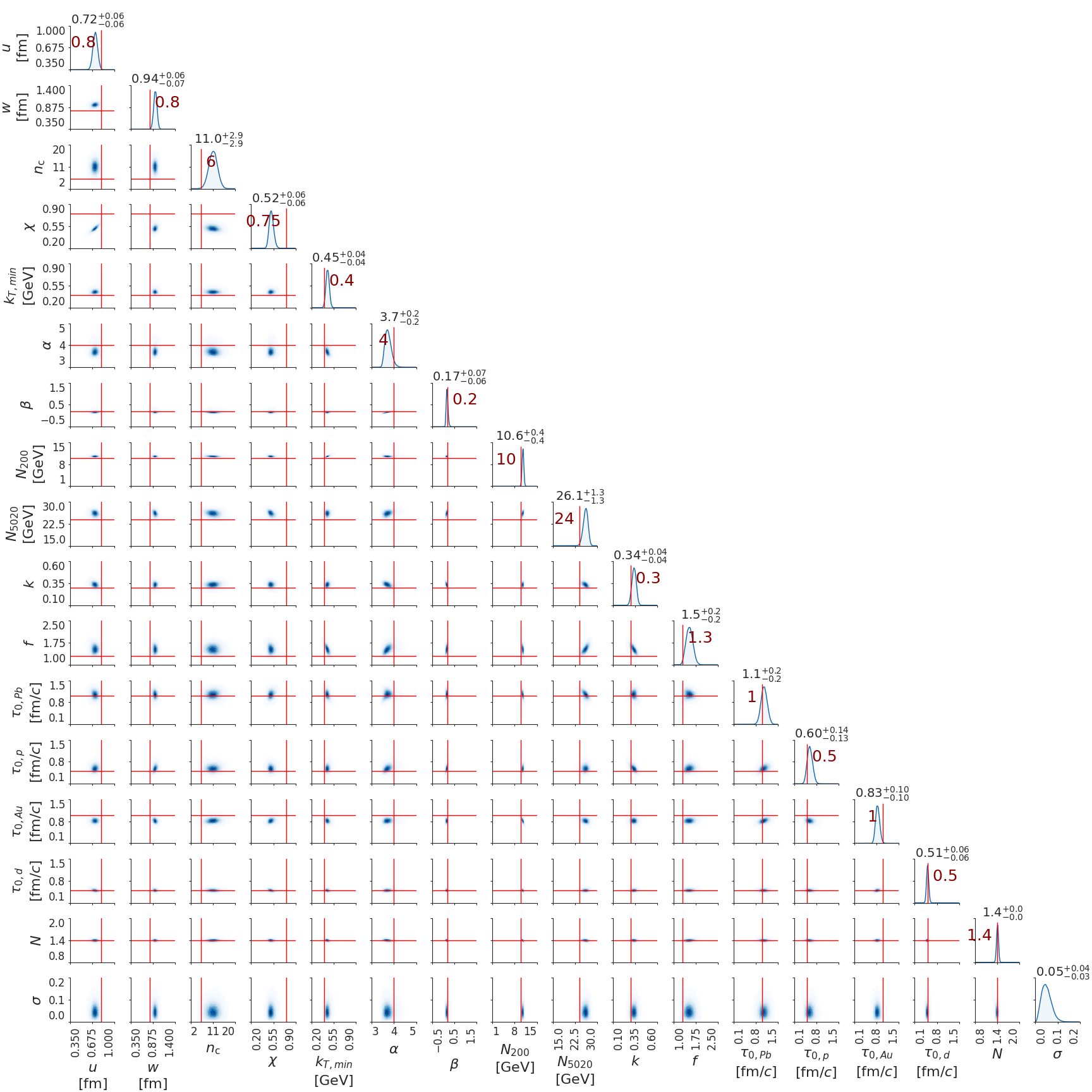}
  \caption
  {
    \label{fig:parampostclosure}
    One- and two-dimensional projections of the parameter posterior distribution for the closure test, a fit to mock data for four collision systems simultaneously.  The red lines indicate the parameter values used to generate the data.
  }
\end{figure*}

The \term{parameter posterior} for the closure test is represented as a ``corner'' or ``triangle'' plot in Fig.~\ref{fig:parampostclosure}.  Visualization of a high-dimensional distribution is inevitably imperfect, but in practice, the one- and two-dimensional projections on the diagonal and lower triangle are often sufficient: to evince any multimodality in the posterior distribution, to reveal correlations among parameters, and to suggest the vicinity of the maximum a posteriori or \term{MAP} point where the posterior is highest.

The diagonal of Fig.~\ref{fig:parampostclosure} exhibits single-parameter posteriors (blue) that generally peak tightly near the parameter values we chose (red).  This indicates both good closure and good constraint, and therefore a successful closure test.  The most prominent deviations are in the $n_c$ and $\chi$ parameters, which both pertain to nucleonic substructure.  Certainly their effects are washed out within the heavy nuclei; we speculate that event averaging, the use of (1+1)D hydrodynamics, and most of all our choice of observable, make $n_c$ and $\chi$ difficult to constrain in general.

The $\sigma$ quantity appearing in the bottom row of Fig.~\ref{fig:parampostclosure} is not a true model parameter but serves as an extra degree of freedom through which the model systematic uncertainty can be estimated~\cite[Ch.~4.4.2]{Bernhard:2018hnz}.  We find that the posterior with respect to this quantity is well in line with expectations and comparable with the artificial uncertainty of the simulated target data.  This suggests that model systematic uncertainty is not a limiting factor in the present analysis.

\begin{figure*}
  \includegraphics[width=0.45 \textwidth]{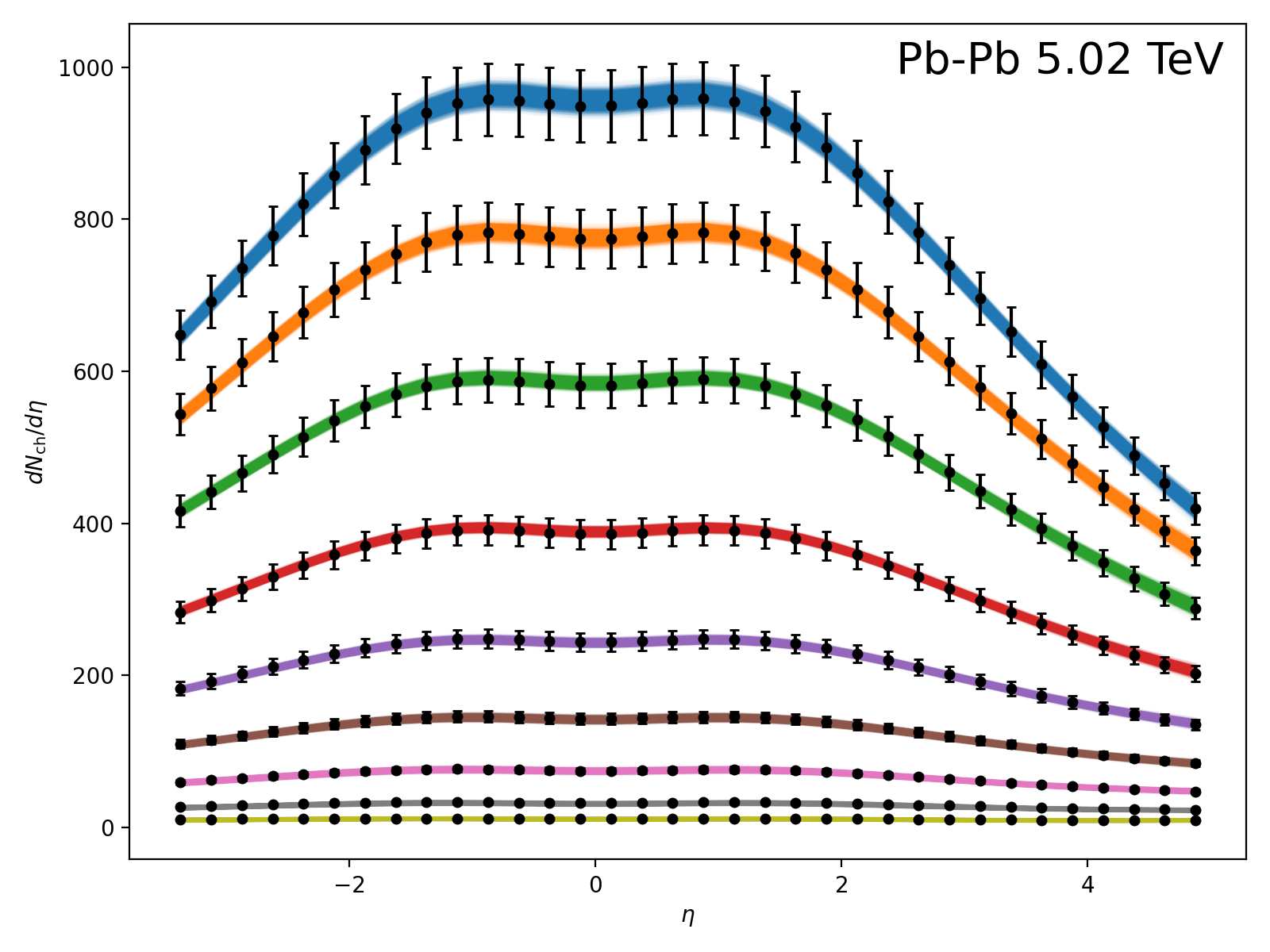}
  \includegraphics[width=0.45 \textwidth]{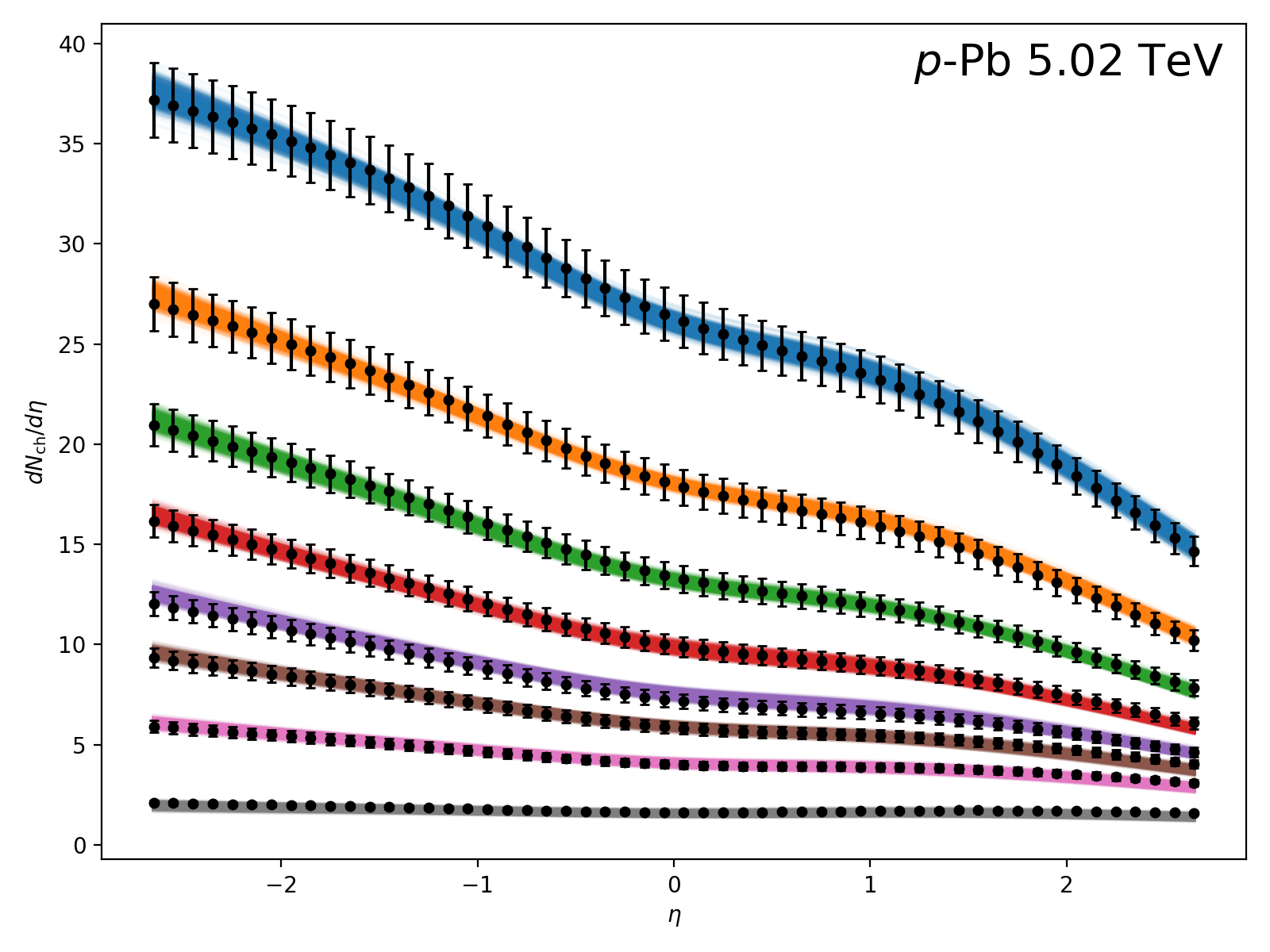} \\
  \includegraphics[width=0.45 \textwidth]{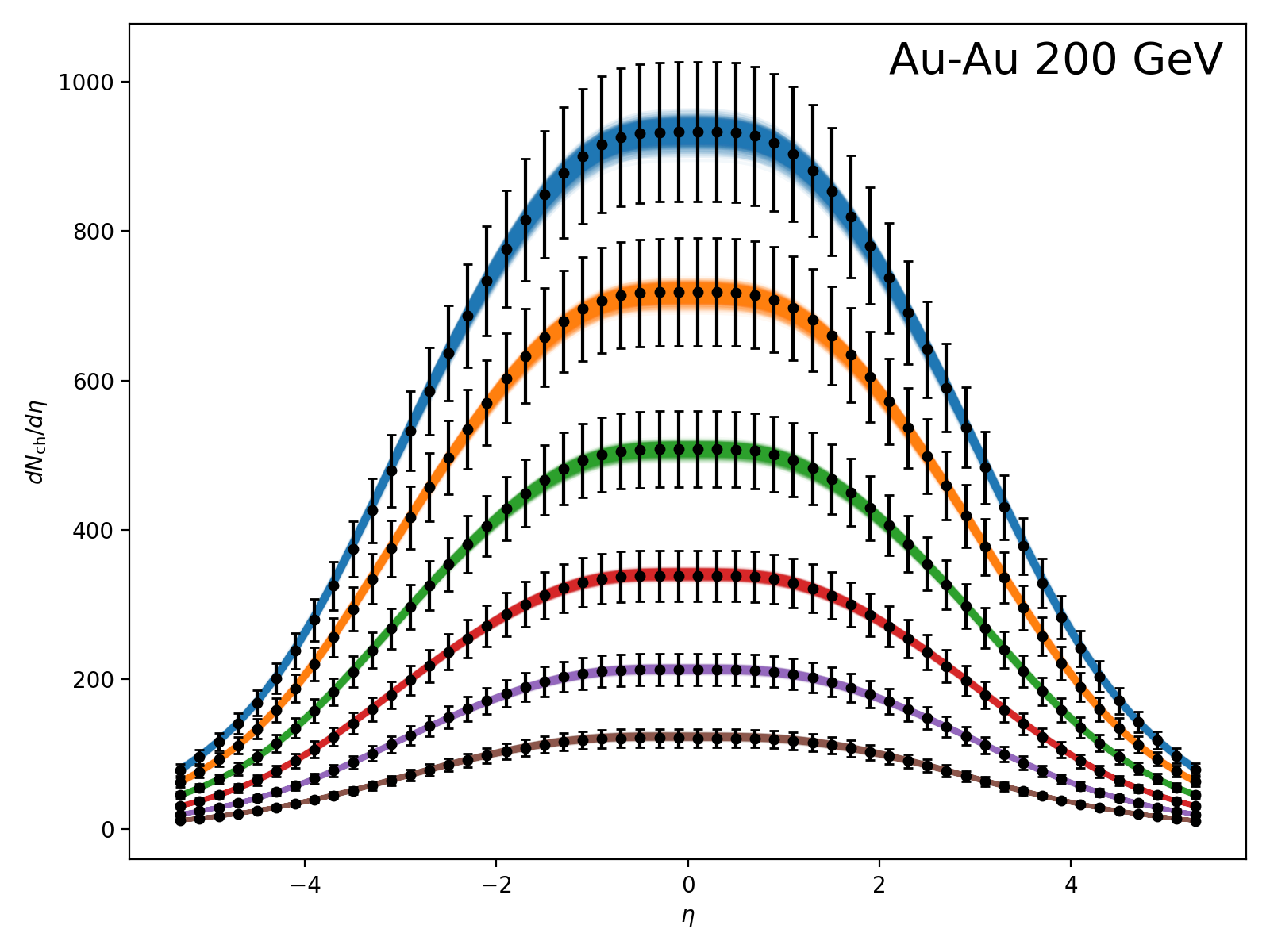}
  \includegraphics[width=0.45 \textwidth]{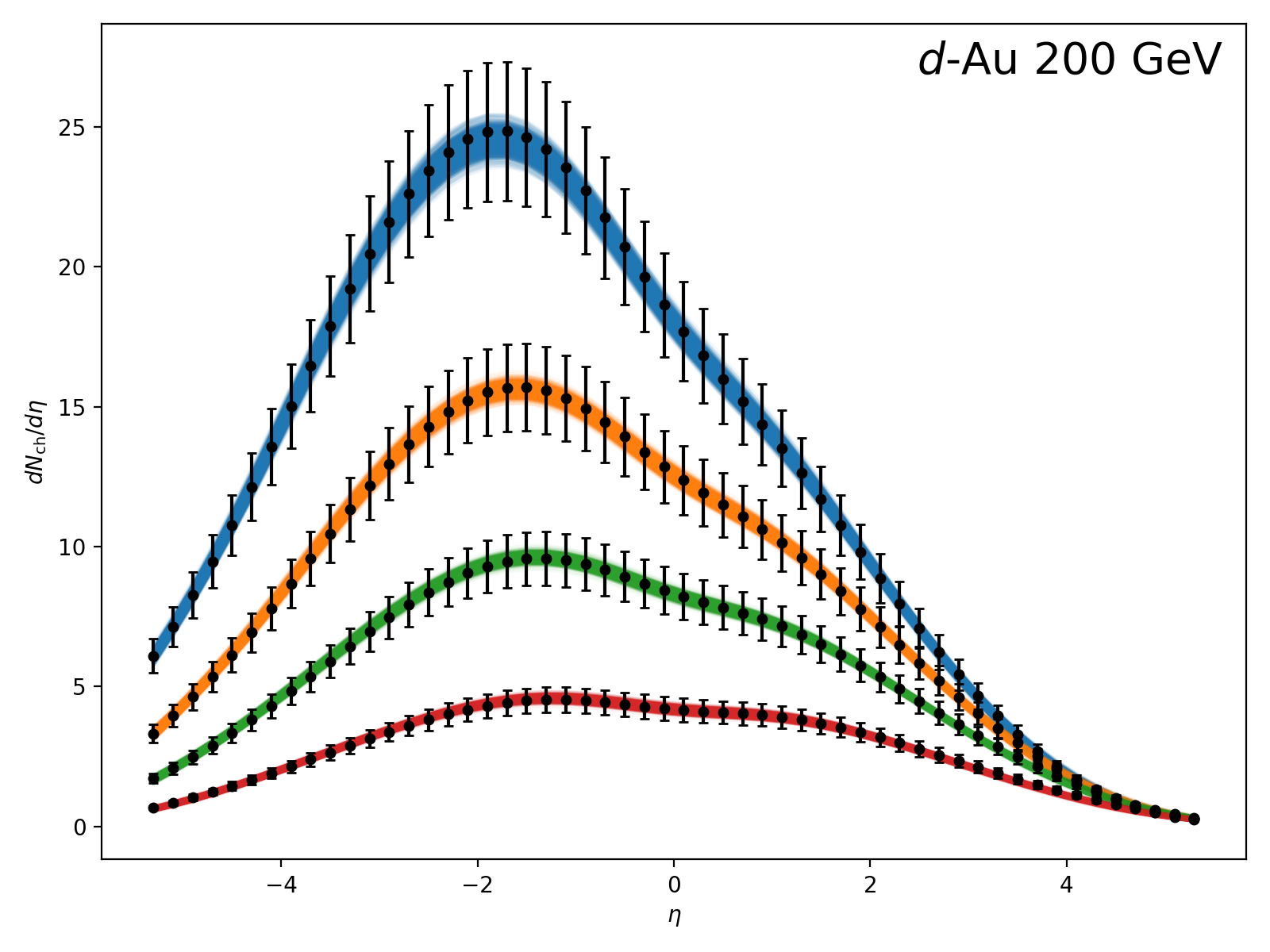}
  \caption
  {
    \label{fig:obspostclosure}
    Observable posteriors from the closure test.  The continuous bands represent 1,000 model predictions of $d\Nch/d\eta$, generated by emulating at parameter points randomly drawn from the posterior.  The colors represent distinct centrality bins per system (see Table~\ref{tab:systems}) and are simply intended as a visual aid.  The data points (black, with error bars) depict the simulated data.
  }
\end{figure*}

The \term{observable posteriors}---plots of observables predicted by the emulator at parameter points sampled from the MCMC chain---are presented for the closure test in Fig.~\ref{fig:obspostclosure}.  Since single-shot hydrodynamics was used to generate the simulated data as well as the training data, the error bars on the simulated data (black) are not statistical uncertainty but simply represent a fraction of the observable value roughly based on experimental uncertainties: 5\% for Pb--Pb and $p$--Pb, 10\% for Au--Au and $d$--Au.  In general, and especially to within uncertainties, the agreement with the simulated data is excellent.  This provides an encouraging validation of the self-consistency of the analysis.

\subsection{Cross-Model Validation}
\label{sec:modeltomodel}

Considering the significant simplifications and approximations in the present evolution model, it is prudent to conduct a second phase of validation, intermediate to the closure test of the previous section and the calibration to experimental data that follows.  In this section, we describe a simulated-data validation in which the target data is generated by a higher-fidelity simulation based on (3+1)D \MUSIC{}~\cite{Schenke:2010nt,Schenke:2010rr,Paquet:2015lta}---a ``cross-model'' comparison.

The present test closely resembles the closure test: we reused the same design and simplified evolution model data to train the emulator, and we targeted data generated using the same parameter point chosen previously.  The only differences are that now the target data is the result of evolving the fully 3D initial conditions with \MUSIC{}'s (3+1)D hydrodynamics and particlization (rather than a 1D profile and the simplified model), and that our goal is to gauge the ability to recover \trentoDDD{} initial-state parameter values despite this difference in evolution models.

\begin{figure*}
  \includegraphics[width=\textwidth]{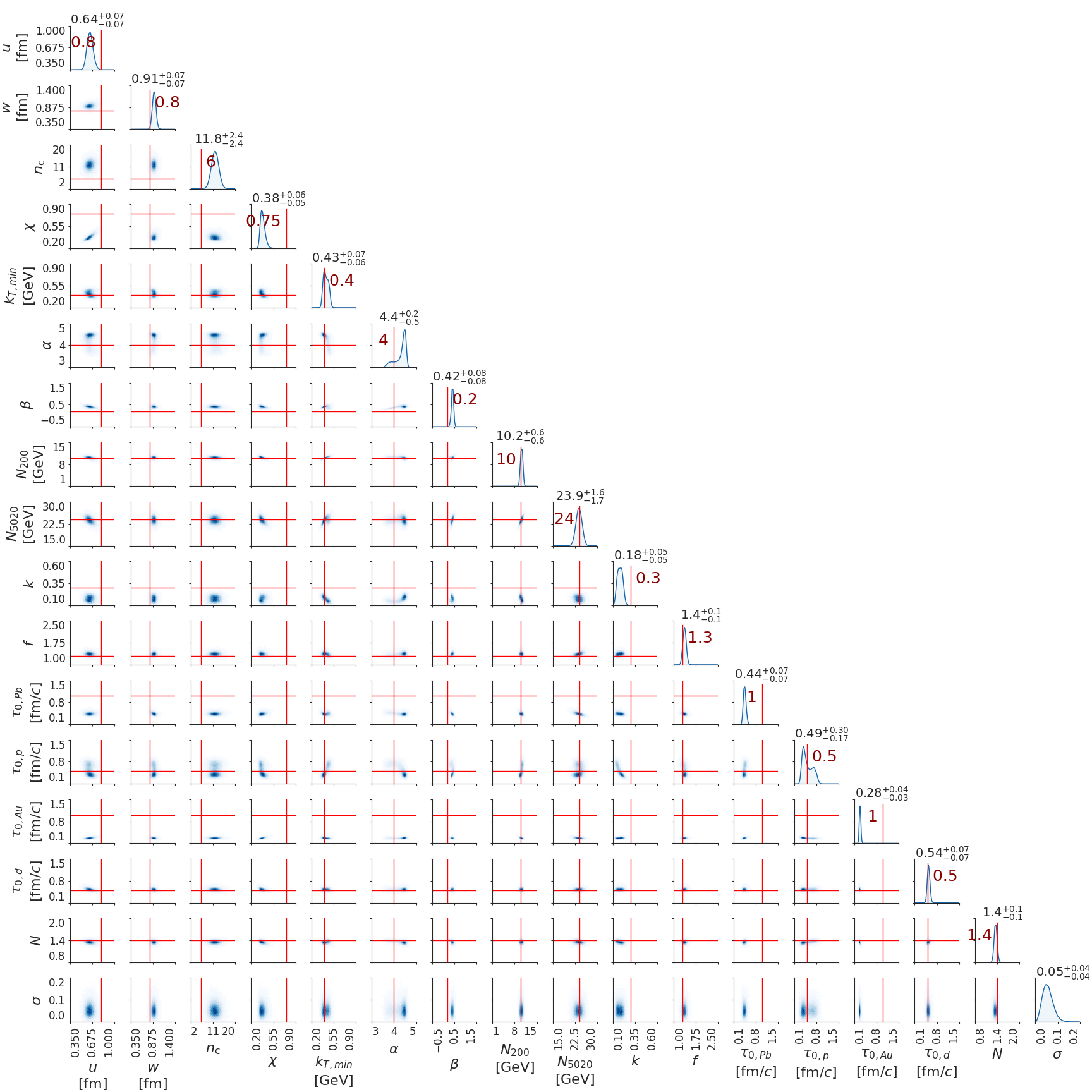}
  \caption
  {
    \label{fig:parampostvsMUSIC}
    Projections of the parameter posterior distribution from the cross-model test, a fit to data generated by (3+1)D \MUSIC{}.
  }
\end{figure*}

\begin{figure*}
  \includegraphics[width=0.45 \textwidth]{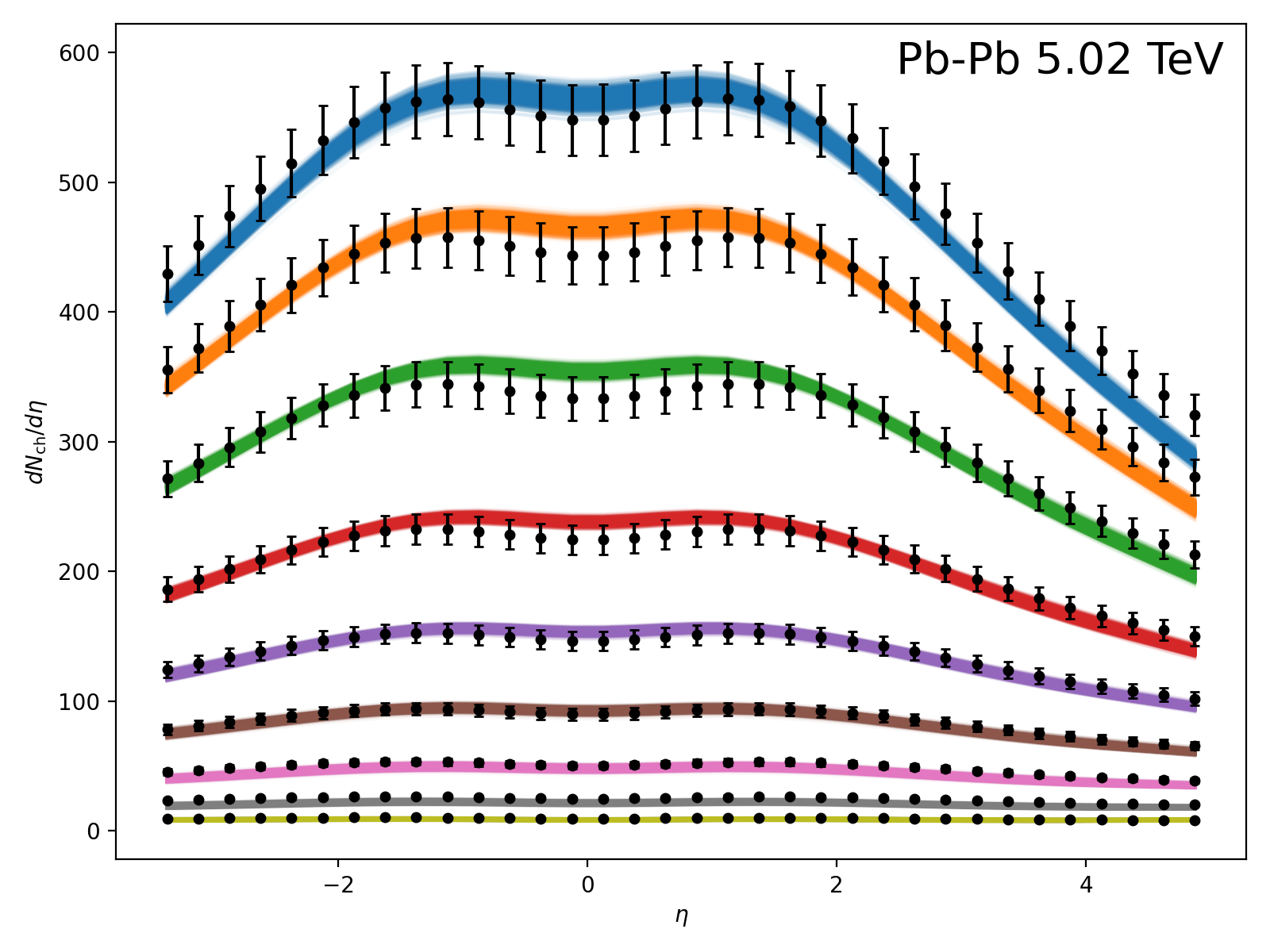}
  \includegraphics[width=0.45 \textwidth]{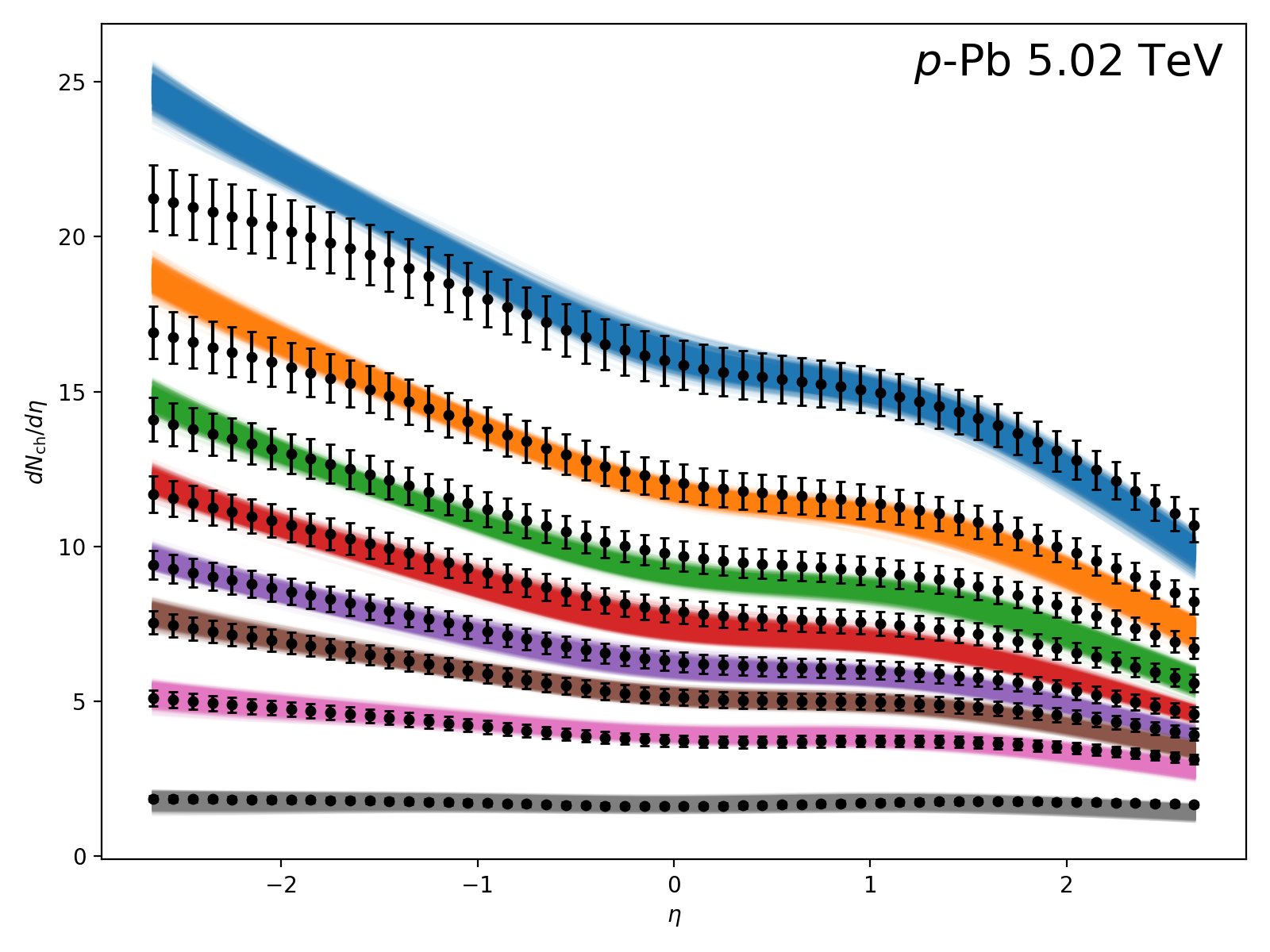} \\
  \includegraphics[width=0.45 \textwidth]{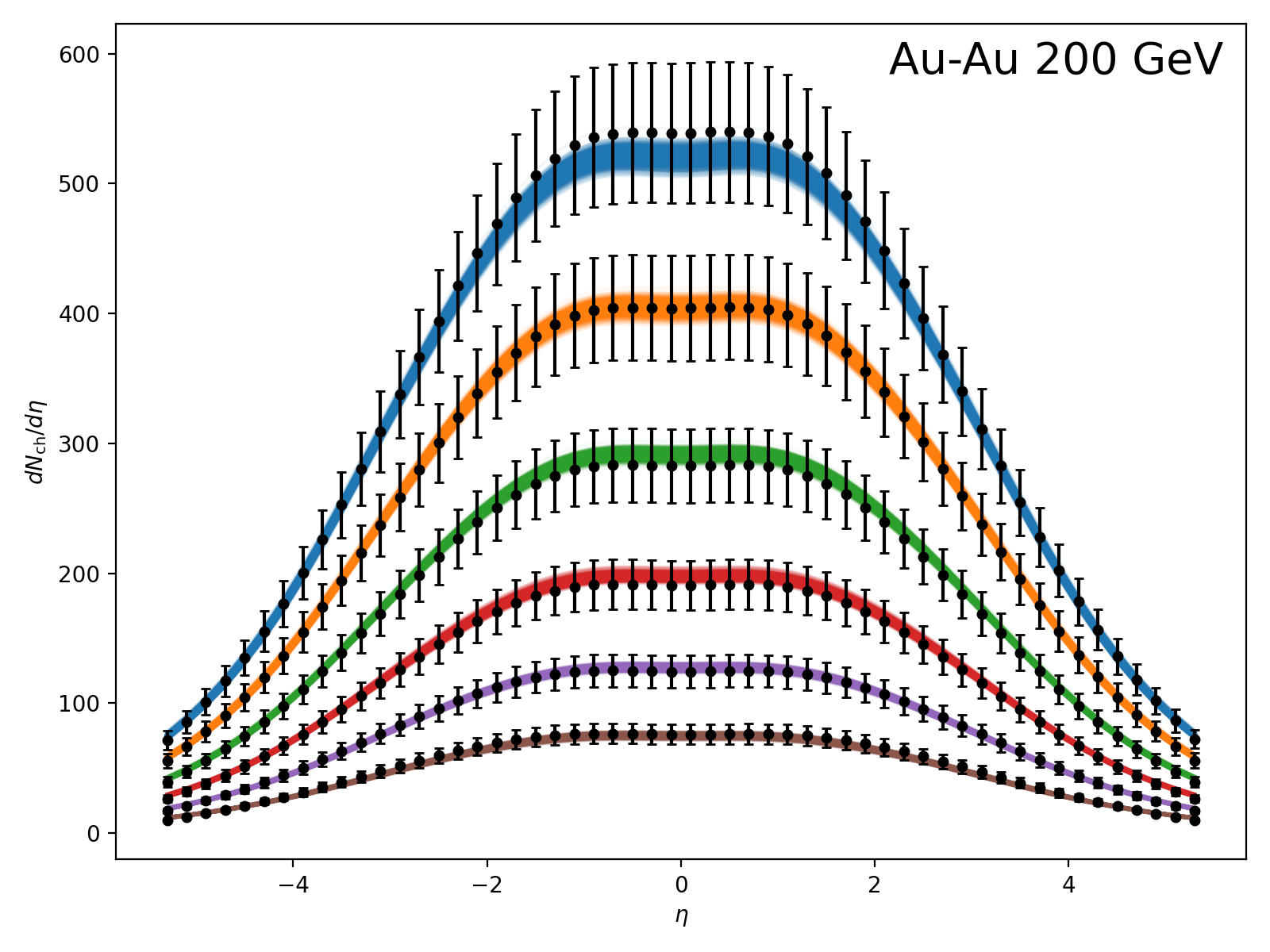}
  \includegraphics[width=0.45 \textwidth]{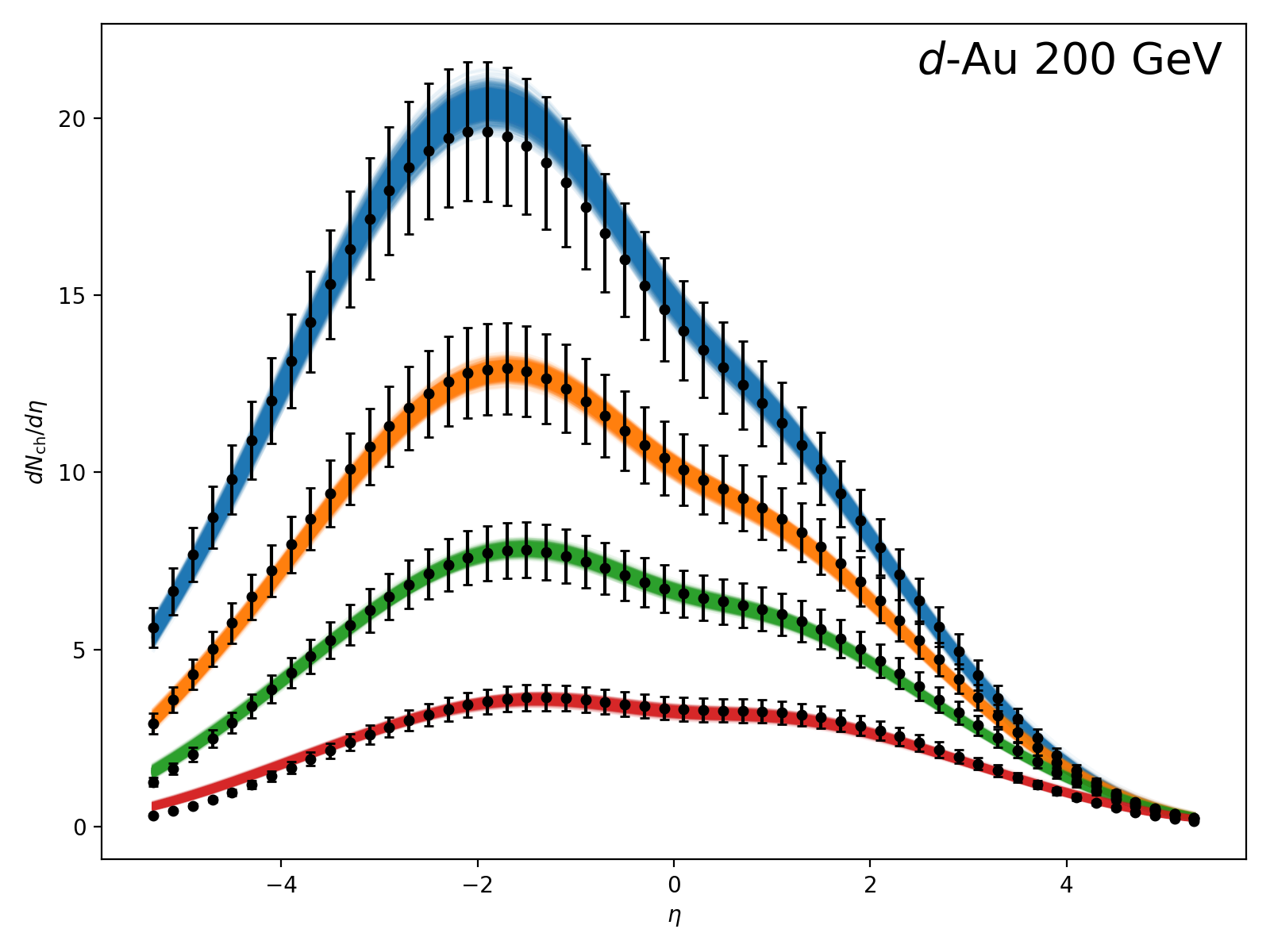}
  \caption
  {
    \label{fig:obspostvsMUSIC}
    Observable posteriors from the cross-model test.  As in Fig.~\ref{fig:parampostclosure}, the colored bands depict $d\Nch/d\eta$ predicted by the emulator (trained on data from the simplified evolution model) for various collision systems and centrality bins.  The data points (black, with error bars) represent the target data, now generated using (3+1)D \MUSIC{}.
  }
\end{figure*}

Figure~\ref{fig:obspostvsMUSIC} presents a clear indication that the analysis was able to describe the simulated data rather well, in an encouraging demonstration of \trentoDDD{}'s expressiveness.  The emulated observable simultaneously matches the system dependence, centrality dependence, and pseudorapidity dependence of the target data to within the latter's assigned uncertainty, at almost all data points except with respect to $p$--Pb.  Comparing with the data points in Fig.~\ref{fig:obspostclosure}, it appears the \MUSIC{}-based calculation predicts markedly lower charged-particle yields than does the LH-based calculation at the same target parameter point, for all systems---an outcome that can be attributed to the differing amounts of work done in the 1D expansion of the simplified hydrodynamics versus the full 3D expansion of \MUSIC{}.

Turning to the parameter posterior, Fig.~\ref{fig:parampostvsMUSIC} shows that the posterior is maximized in a region compatible with many of the chosen parameter values, though understandably with lower accuracy than the closure test (Fig.~\ref{fig:parampostclosure}), since we are now calibrating on data generated by a far more sophisticated evolution model. While the more sophisticated hydrodynamic evolution and particlization stages affect the values of the extracted the initial-conditions parameters, the general features of the chosen initial conditions are still recoverable.  As in the closure test, the nuclear configuration-related parameters were difficult to constrain, presumably because we are not currently calibrating on observables known to be sensitive to the nuclear configuration such as elliptic and triangular flow.  However, most of the parameters governing the shape of the charged-hadron distribution---$\kTmin$, $\beta$, $N_{200}$, $N_{5020}$, and $\flatness$---\emph{were} accurately determined in the cross-model test, indicating that the spectral shape predicted by a high-fidelity evolution model is generally well-reproduced by the simplified model.  The most prominent difference between the models appears to be a matter of scale, and this is reflected primarily by the discrepancies seen in the $\tau_0$ family of parameters, where lower values of $\tau_0$ were favored.  The reason is that a smaller $\tau_0$ allows the linearized hydro to do more work and thus compensate for the difference in the amount of work done by the expanding 3D hydro vs. 1D hydro and therefore an emulator trained on the simplified evolution model achieves a better fit to the (lower-multiplicity) \MUSIC{} data given a smaller value of $\tau_0$ than was supplied to \MUSIC{}.

\subsection{Comparison to Data}
\label{sec:results}

We now compare the results of our simplified evolution model to data taken at RHIC and LHC.  As in preceding sections, we perform a Bayesian analysis to estimate the parameter values that cause simulation to best reproduce data.  However, unlike in the cross-model validation and particularly the closure test, there is no a priori guarantee that \emph{any} point in parameter space describes the data from experiment.  Of course, \emph{some} point will offer a better fit to data than the rest; as before, this is termed the MAP point, and the goodness of the fit it produces will suggest the adequacy of the overall setup: models, parametrizations, prior, design, observables, emulator, MCMC, data, and uncertainties.

For the model-to-data comparison, we reuse every component of the earlier tests except the target data, which is now taken from the experimental results referenced in Table~\ref{tab:systems}.  This reduces opportunities for mistakes and ensures that the earlier tests remain as relevant as possible to the final analysis.

\begin{figure*}
  \includegraphics[width=\textwidth]{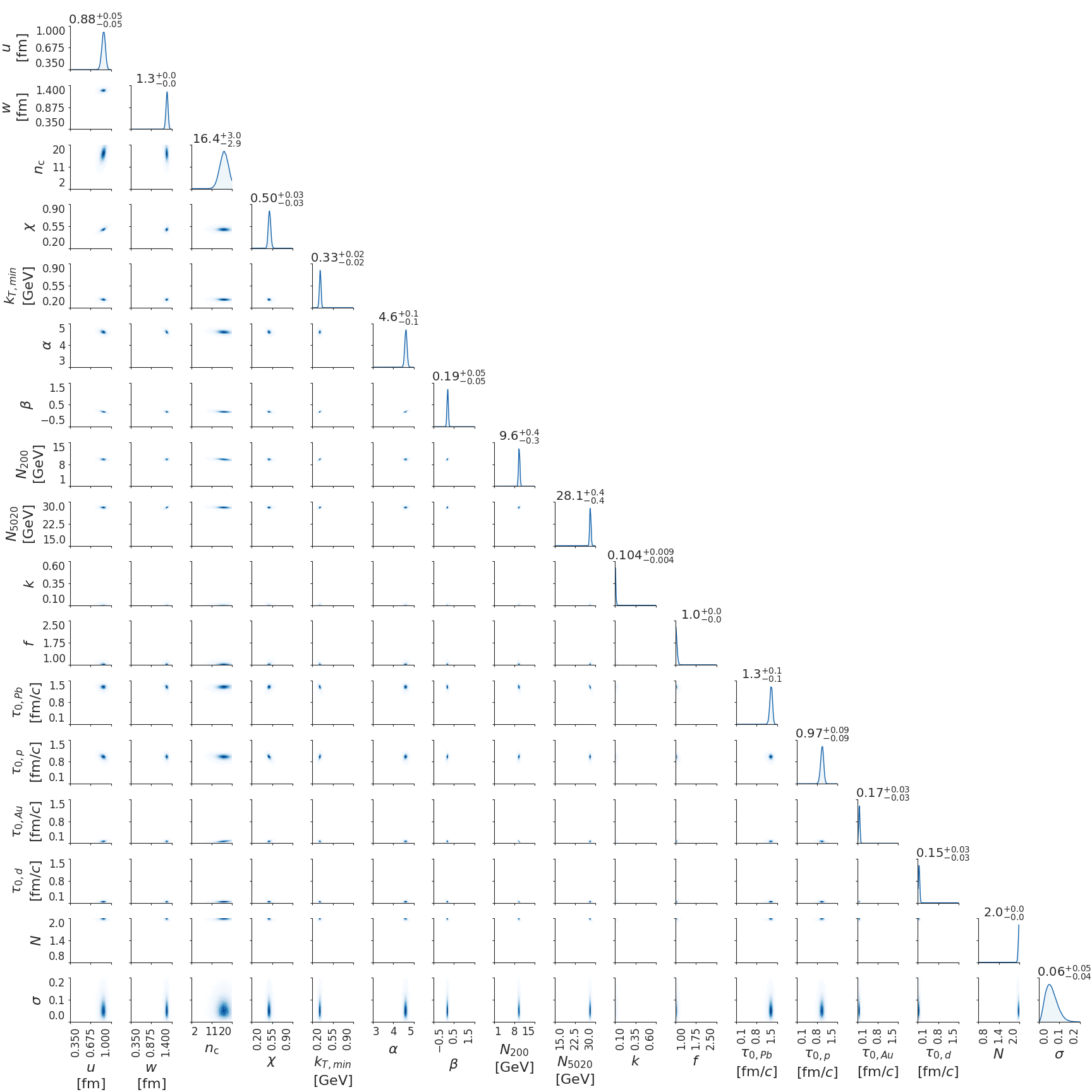}
  \caption
  {
    \label{fig:parampost}
    One- and two-dimensional projections of the parameter posterior distribution from a simultaneous fit to experimental data for four collision systems (Table~\ref{tab:systems}).
  }
\end{figure*}

The posterior corner plot for the model-to-data comparison is shown in Fig.~\ref{fig:parampost}.  There is excellent constraint in all parameters, meaning the posterior has a sharp and apparently unambiguous maximum.  However, the maximum appears to be at or near the edge of the prior along some parameter axes, which is not optimal because computing the posterior in this region almost certainly involves emulator extrapolation, rather than the interpolation to which GPs are better suited, and because the posterior is effectively being truncated by the artificially discontinuous prior.  As the prior was generally chosen to encompass all sensible parameter values, expanding it in the directions suggested by Fig.~\ref{fig:parampost} is not straightforward.

\begin{figure*}
  \includegraphics[width=0.45 \textwidth]{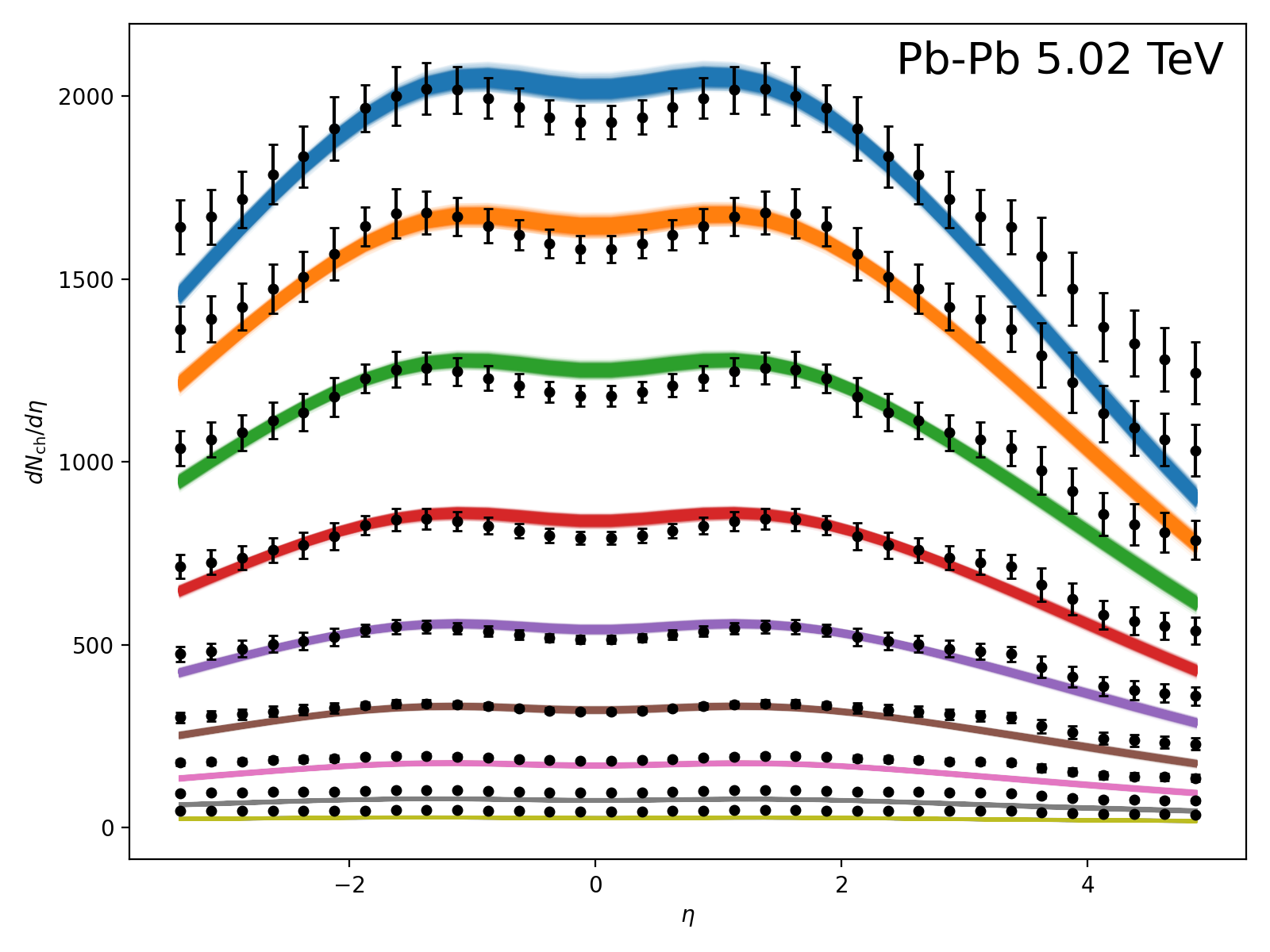}
  \includegraphics[width=0.45 \textwidth]{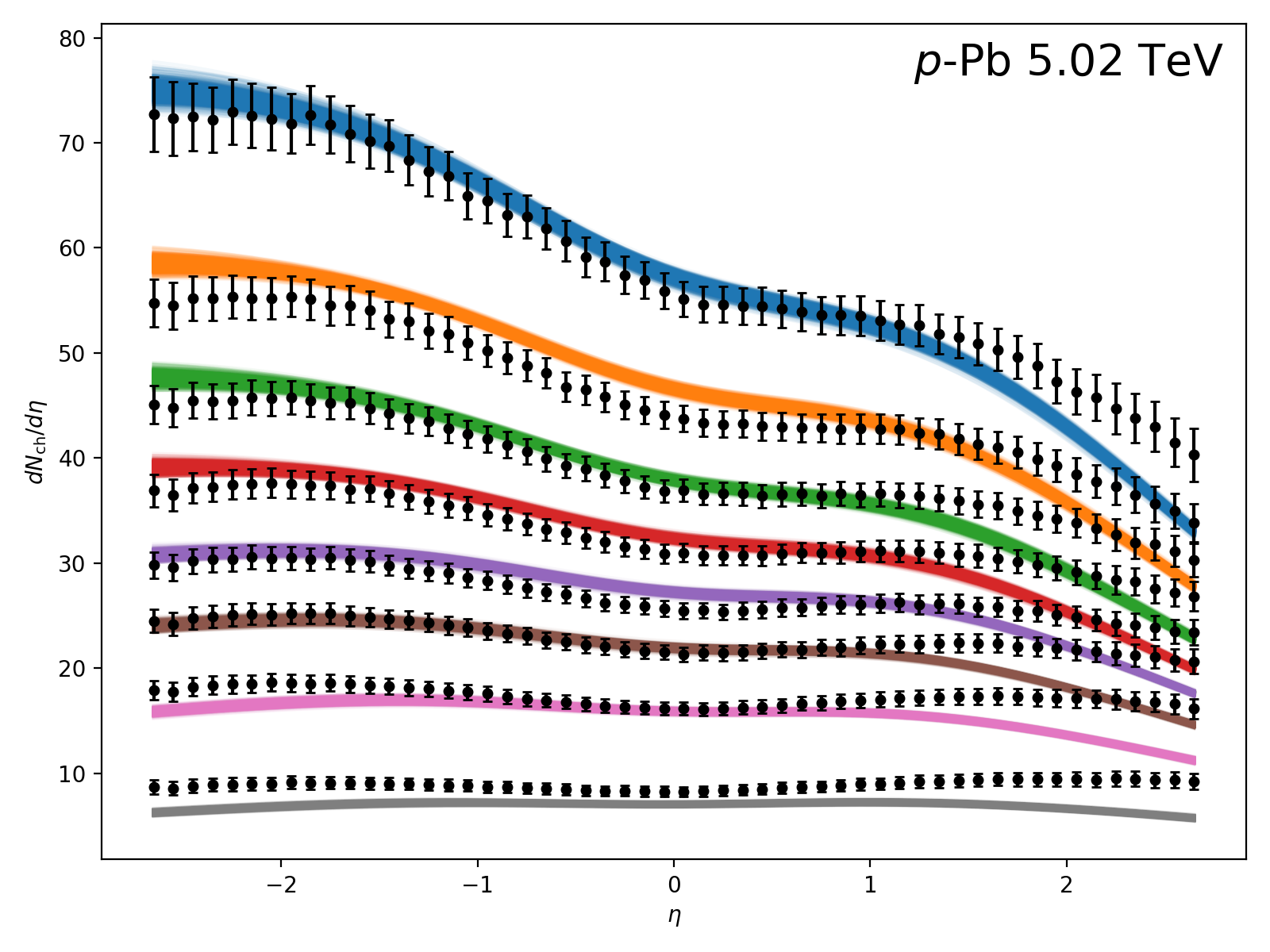} \\
  \includegraphics[width=0.45 \textwidth]{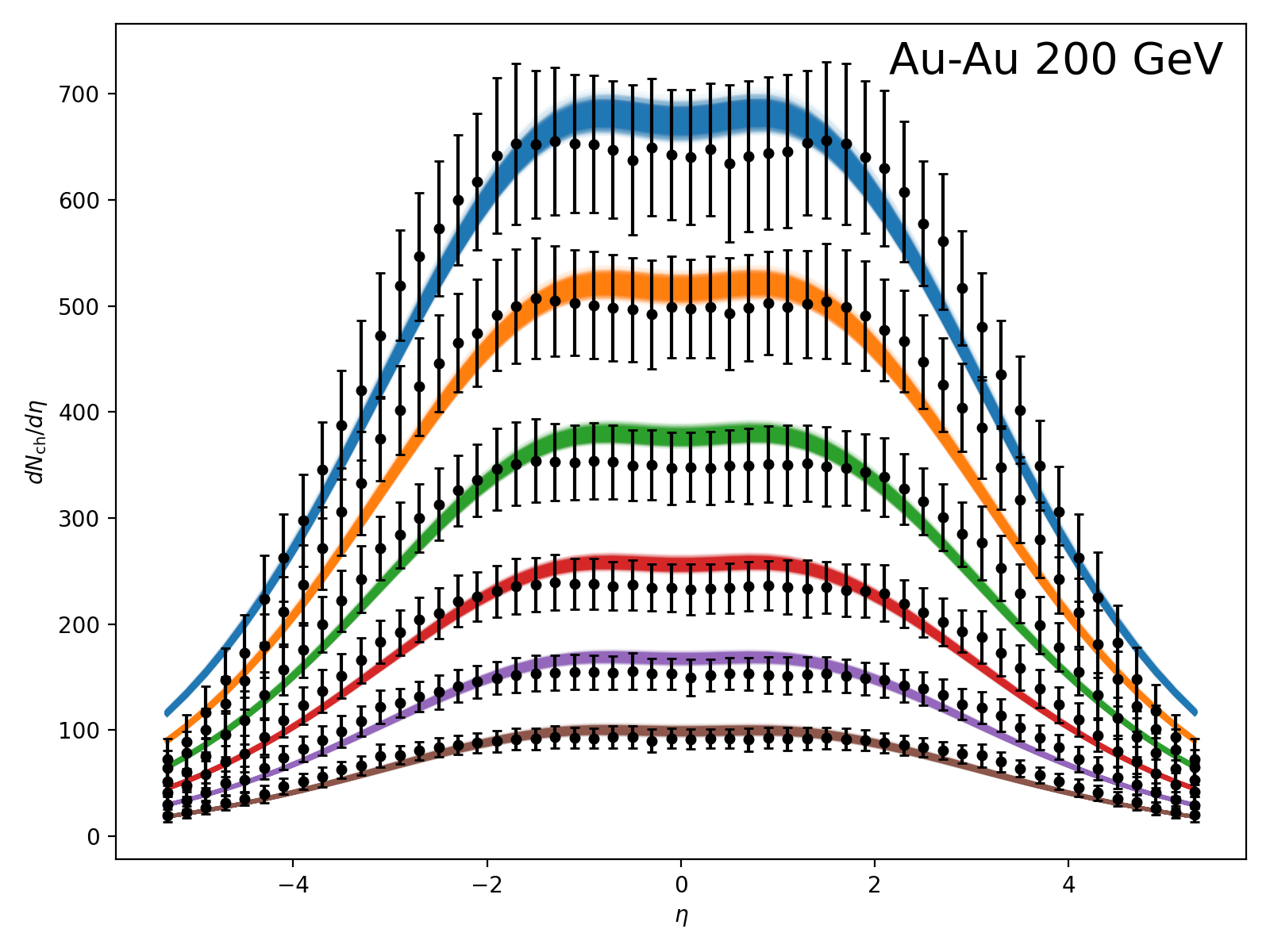}
  \includegraphics[width=0.45 \textwidth]{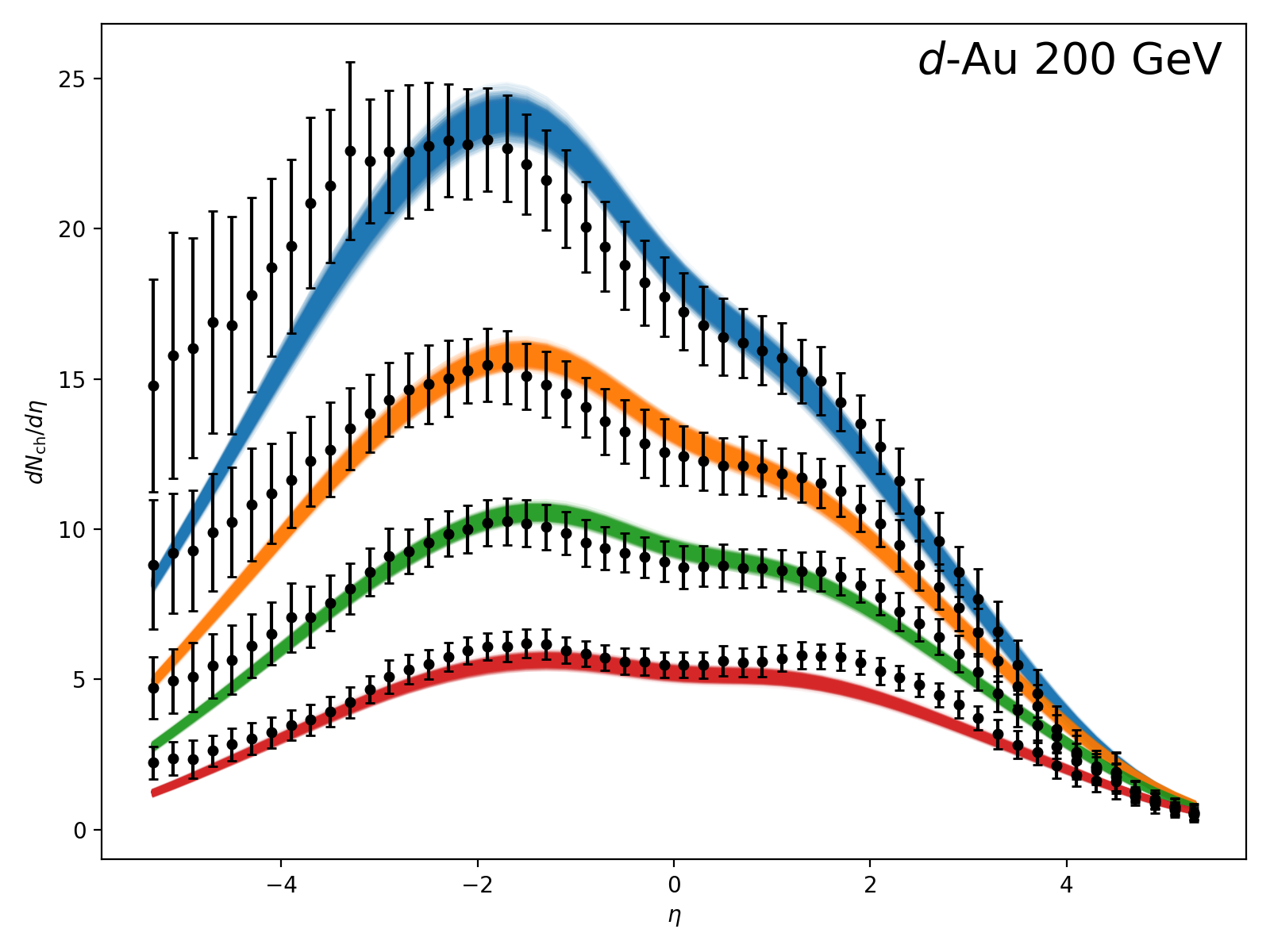}
  \caption
  {
    \label{fig:obspost}
    Observable posteriors from the simultaneous fit to experimental data for four collision systems.  The continuous bands represent 1,000 model predictions of $d\Nch/d\eta$, generated by emulating at parameter points randomly drawn from the posterior.  The colors represent distinct centrality bins per system and are simply intended as a visual aid.  The data points (black, with error bars) are from ALICE~\cite{ALICE:2016fbt} (Pb--Pb), ATLAS~\cite{ATLAS:2015hkr} ($p$--Pb), and PHOBOS~\cite{Back:2002wb} (Au--Au), \cite{PHOBOS:2004fzb}~($d$--Au).  Errors have been symmetrized by taking the larger of the positive and negative values for each contribution (systematic and statistical, when given) and adding the contributions in quadrature.
  }
\end{figure*}

Figure~\ref{fig:obspost} presents the observable posteriors for the model-to-data comparison with the target experimental data superimposed.  Within uncertainties, there is generally good agreement with the symmetric heavy-ion data (Pb--Pb and Au--Au) and reasonable agreement with the small-systems data ($p$--Pb and $d$--Au), in both pseudorapidity and centrality. Overall we find that \trentoDDD{} with our simplified evolution model is well capable of describing our reference data, thus passing our final test regarding its viability as an initial-conditions model for ultrarelativistic nucleus--nucleus and proton--nucleus collisions.

\section{Summary and Outlook}

We have extended the well-studied midrapidity \trento{} initial-conditions model to three dimensions, in order to provide a fast and flexible parametric initial-conditions model for ultrarelativsitic heavy-ion collisions in full 3D, suitable for use with state-of-the-art (3+1)D relativistic viscous hydrodynamics models.  \trentoDDD{} builds on \trento{}'s success at modeling the configurations and interacting thicknesses of colliding nuclei, longitudinally extending the scope of the model to a central fireball near midrapidity plus two fragmentation regions at forward and backward rapidities.  We have demonstrated \trentoDDD{}'s viability through a series of tests that combine \trentoDDD{} with linearized (1+1)D hydrodynamic evolution and a cylindrically-symmetric particlization model, in order to account for the broadening of the final-state charged-particle rapidity distributions due to the expansion of the QGP. Utilizing this simplified evolution model, we first performed a closure test to demonstrate the validity of our analysis framework.  We subsequently progressed to a large-scale calibration of the simplified evolution model, utilizing as mock data the rapidity distributions of charged hadrons generated by a high-fidelity (3+1)D relativistic viscous hydrodynamics model with \trentoDDD{} initial conditions.  This calibration demonstrated the ability of our simplified evolution model to act as a stand-in for a high-fidelity model and still allow recovery of salient features of the (known) \trentoDDD{} initial state.  In a final test, we calibrated our \trentoDDD{} initial state and simplified evolution model to actual experimental data, to demonstrate \trentoDDD{}'s ability to provide initial conditions that can result in realistic final-state distributions.

Moving forward, the next step will be to perform a calibration to rapidity-dependent RHIC and LHC data using \trentoDDD{} with the high-fidelity hydrodynamics model, rather than with the simplified hydrodynamics model of the present work.  Utilizing such a state-of-the-art model with data from asymmetric and/or small collision systems, and data away from midrapidity, should allow for an improved extraction of QGP and initial-state properties and for a better disambiguation between the QGP and initial-state characteristics.

\begin{acknowledgments}
This work was supported by the U.S. Department of Energy Grant No. DE-FG02-05ER41367 (SAB, JFP and DS).
DS is also supported by NSF Grant OAC-1550225. WK is supported by the Laboratory Directed Research and Development Program (LDRD) at Los Alamos National Laboratory. This research used resources of the National Energy Research Scientific Computing Center (NERSC), a U.S. Department of Energy Office of Science User Facility operated under Contract No. DE-AC02-05CH11231.
\end{acknowledgments}

\newpage

\appendix
\section{Linearized Hydrodynamics}
\label{app:LH}

In this appendix we present a complete derivation of the simplified hydrodynamics model introduced in Sec.~\ref{sec:LH}.  For further reading on (1+1)D ideal hydrodynamics in heavy-ion collisions, we suggest Refs.~\cite{Satarov:2006jq,Beuf:2008vd,Bialas:2010ax,Stephanov:2014hfa}.

We start with the continuity equation,
\begin{equation}
\nabla_\mu \Tmunu = J^\nu\text{,} \label{eq:LHcont}
\end{equation}
where $J^\nu$ is a source term (formulated below) accounting for the sudden appearance of the initial energy distribution.  Note that we have used the covariant derivative $\nabla_\mu$, rather than the partial derivative $\partial_\mu$, since we are working in Milne ($\tau$--$\etas$) coordinates.  Expanding $\nabla_\mu$ yields
\begin{equation}
\partial_\mu \Tmunu + \Gamma^\mu_{\mu\lambda} T^{\lambda\nu} + \Gamma^\nu_{\mu\lambda} T^{\mu\lambda} = J^\nu\text{,} \label{eq:LHcont2}
\end{equation}
with $\Gamma^\gamma_{\alpha\beta}$ being Christoffel symbols of the second kind.  For the metric
\begin{equation}
g_{\mu\nu} = \begin{pmatrix}
1 & 0 \\
0 & -\tau^2
\end{pmatrix}, \quad
g^{\mu\nu} = \begin{pmatrix}
1 & 0 \\
0 & -1/\tau^2
\end{pmatrix}\text{,} \label{eq:LHg}
\end{equation}
the Christoffel symbols are:
\begin{align}
\Gamma^0_{\alpha\beta} & = \begin{pmatrix}
0 & 0 \\
0 & \tau
\end{pmatrix} \\
\Gamma^3_{\alpha\beta} & = \begin{pmatrix}
0 & 1/\tau \\
1/\tau & 0
\end{pmatrix}\text{.}
\end{align}
(Note that we are using $0$ and $3$ to represent the $\tau$ and $\etas$ coordinates respectively, skipping $1$ and $2$ since there are no transverse dimensions.)  Given the above (and $\Tmunu = T^{\nu\mu}$), Eq.~\eqref{eq:LHcont2} becomes
\begin{align}
\left(\partial_0 + \frac{1}{\tau}\right) T^{00} + \partial_3 T^{03} + \tau T^{33} & = J^0 \label{eq:LHdiff0} \\
\left(\partial_0 + \frac{3}{\tau}\right) T^{03} + \partial_3 T^{33} & = J^3\text{.} \label{eq:LHdiff3}
\end{align}

Before plugging in the approximate form \eqref{eq:LHTmunu} for $\Tmunu$, it is convenient to define
\begin{equation}
\LHpz \equiv \left(1 + \cssq\right) \LHe ~ \LHvz\text{,}
\end{equation}
on the basis of $T^{0i}$ representing momentum density in the definition of the energy--momentum tensor.  Thus we obtain the much more compact form
\begin{equation}
\Tmunu = \begin{pmatrix}
\LHe & \frac{1}{\tau} \LHpz \\
\frac{1}{\tau} \LHpz & \frac{c_s^2}{\tau^2} \LHe
\end{pmatrix}\text{,}
\end{equation}
and the differential equations \eqref{eq:LHdiff0} and \eqref{eq:LHdiff3} expand as follows:
\begin{align}
\partial_0 \LHe + \frac{1 + \cssq}{\tau} \LHe + \frac{1}{\tau} \partial_3 \LHpz & = J^0 \label{eq:LHdiff0pz} \\
\frac{1}{\tau} \partial_0 \LHpz + \frac{2}{\tau^2} \LHpz + \frac{\cssq}{\tau^2} \partial_3 \LHe & = J^3 \label{eq:LHdiff3pz}
\end{align}

To remove the factors of $\tau$, we change variables to a new time
\begin{equation}
\logtime \equiv \ln \left( \frac{\tau}{\tau_0} \right)
\end{equation}
with $\partial_0 \to \pdlogtime / \tau$ such that all terms have equivalent powers of $\tau$ that can then be factored out.  In terms of $\logtime$, Eqs.~\eqref{eq:LHdiff0pz} and \eqref{eq:LHdiff3pz} become
\begin{align}
\pdlogtime \LHe + \left(1 + \cssq\right) \LHe + \partial_3 \LHpz & = \tau J^0 \label{eq:LHdiff0t} \\
\cssq \partial_3 \LHe + \pdlogtime \LHpz + 2 \LHpz & = \tau^2 J^3\text{.} \label{eq:LHdiff3t}
\end{align}
At this point the derivation will benefit from specifying $J^\nu$.  Suppose that in the vicinity of $\tau \to \tau_0$, the energy density resembles a discontinuous jump to the initial conditions (i.e., $\Theta(\tau - \tau_0) ~ \epsIC(\etas)$, using the Heaviside theta), and suppose also that the initial flow is zero.  We then obtain for the source term
\begin{align}
J^0 & = \delta(\tau - \tau_0) \epsIC(\etas) \\
J^3 & = 0\text{.}
\end{align}
$J^0$ transforms to the new time coordinate as follows:
\begin{align}
J^0 & = \delta(\tau - \tau_0) \epsIC(\etas) \\
& = \delta\left\lbrace \tau_0 \left( e^{\logtime} - 1 \right) \right\rbrace \epsIC(\etas) \\
& \approx \frac{1}{\tau_0} \delta(\logtime) \epsIC(\etas)\text{.}
\end{align}
For simplicity, we will write the right side of Eq.~\eqref{eq:LHdiff0t} as
\begin{align}
f(t, \etas) & \equiv \frac{\tau}{\tau_0} \delta(\logtime) \epsIC(\etas) \\
& = e^{\logtime} \delta(\logtime) \epsIC(\etas)\text{.}
\end{align}

With $J^3$ nullified, it becomes simpler to combine Eqs.~\eqref{eq:LHdiff0t} and~\eqref{eq:LHdiff3t} into one equation for each of $\LHe$ and $\LHpz$.  Doing so, we obtain
\begin{align}
\label{eq:LHtelee}
\LL \LHe & = -\left( \pdlogtime + 2 \right) f \\
\label{eq:LHtelepz}
\LL \LHpz & = \cssq \partial_3 f\text{,}
\end{align}
with
\begin{equation}
\LL \equiv -\pdlogtime^2 - \left(3 + \cssq\right) \pdlogtime - 2\left(1 + \cssq\right) + \cssq \partial_3^2
\end{equation}
a differential operator common to both equations.  (We mention in passing that $\LL$ has the general form of the damped telegrapher's equation or Khalatnikov equation, although we were unable to locate in the literature an extant solution useful for the present, specific case.)

Since Eqs.~\eqref{eq:LHtelee} and~\eqref{eq:LHtelepz} are inhomogeneous, our strategy will be to apply the Green's function method.  We proceed in Fourier space; explicitly, the Fourier transform $\FF$ over both coordinates is defined (for some real-space function $\psi$)
\begin{equation}
\FF\lbrace \psi \rbrace = \frac{1}{2 \pi} \int_{-\infty}^{+\infty} d\logtime \int_{-\infty}^{+\infty} d\etas ~ e^{-i \omega \logtime + i \kappa \etas} ~ \psi(\logtime, \etas)\text{.}
\end{equation}
Accordingly, the derivatives in $\LL$ transform as
\begin{align}
\pdlogtime & \to i \omega \\
\pdetas & \to -i \kappa\text{,}
\end{align}
and therefore $\LL$ transforms to
\begin{equation}
\LLF \equiv \FF\lbrace \LL \rbrace = \omega^2 - \left( 3 + \cssq \right) i \omega - 2 \left( 1 + \cssq \right) - \cssq \kappa^2\text{.}
\end{equation}
Factoring $\LLF$ in terms of $\omega$ yields
\begin{equation}
\LLF = \left( \omega - \omega_+(\kappa) \right) \left( \omega - \omega_-(\kappa) \right)\text{,}
\end{equation}
with
\begin{equation}
\omega_\pm(\kappa) \equiv -\frac{3 + \cssq}{2 i} \pm \Omega(\kappa)
\end{equation}
and
\begin{equation}
\Omega(\kappa) \equiv \sqrt{ \cssq \kappa^2 - \frac{\left( 1 - \cssq \right)^2}{4} }\text{.}
\end{equation}
Equations~\eqref{eq:LHtelee} and~\eqref{eq:LHtelepz} then transform to
\begin{align}
\LLF E & = -\left( i \omega + 2 \right) F \\
\LLF P & = -i \cssq \kappa F\text{,}
\end{align}
where capital letters represent the functions in Fourier space, viz., $E(\omega, \kappa) = \FF\left\lbrace \LHe(\logtime, \etas) \right\rbrace$, $P(\omega, \kappa) = \FF\left\lbrace \LHpz(\logtime, \etas) \right\rbrace$, and $F(\omega, \kappa) = \FF\left\lbrace f(\logtime, \etas) \right\rbrace$.

The respective Fourier-space Green's functions $G_\eps$ and $G_p$ are
\begin{align}
G_\eps(\logtime', \etas'; \omega, \kappa) & = -\frac{i \omega + 2}{\LLF} \FF\left\lbrace \delta(\logtime - \logtime') \delta(\etas - \etas') \right\rbrace \\
& = -\frac{i \omega + 2}{2 \pi ~ \LLF} ~ e^{-i \omega \logtime' + i \kappa \etas'} \\
G_p(\logtime', \etas'; \omega, \kappa) & = -\frac{i \cssq \kappa}{2 \pi ~ \LLF} ~ e^{-i \omega \logtime' + i \kappa \etas'}\text{,}
\end{align}
with the primed coordinates being those over which convolution will be performed, and the unprimed coordinates specifying where the solution is to be evaluated.  For the present purpose it suffices to evaluate the solution numerically, so we will not attempt to fully inverse-transform the Green's functions to analytic expressions.  However, the $\omega \to \logtime$ inverse transform \emph{can} be computed analytically, and doing so will make the numerical evaluation more efficient, so this is our next step.  For $G_\eps$:
\begin{align}
\FF_\omega^{-1}\left\lbrace G_\eps \right\rbrace & = -\frac{1}{\sqrt{2 \pi}} e^{i \kappa \etas'} \int_{-\infty}^{+\infty} d\omega ~ \times \nonumber \\
& \qquad \qquad \frac{ e^{i \omega \Delta\logtime} \left( i \omega + 2 \right) }{ \left( \omega - \omega_+ \right) \left( \omega - \omega_- \right) } \\
& = \sqrt{2 \pi} ~ e^{i \kappa \etas'} ~ e^{ -\frac{3 + \cssq}{2} \Delta\logtime } ~ \times \nonumber \\
& \qquad \qquad \left[ \cos\left( \Omega \Delta\logtime\right) + \frac{1 - \cssq}{2 \Omega} \sin\left( \Omega \Delta\logtime \right) \right]\text{,}
\end{align}
with $\Delta\logtime \equiv \logtime - \logtime'$, and $\FF_\omega^{-1}$ signifying the one-dimensional ($\omega \to \logtime$) inverse Fourier transform.  Likewise, for $G_p$:
\begin{align}
\FF_\omega^{-1}\left\lbrace G_p \right\rbrace & = -\frac{i \cssq \kappa}{\sqrt{2 \pi}} e^{i \kappa \etas'} \int_{-\infty}^{+\infty} d\omega ~ \times \nonumber \\
& \qquad \qquad \frac{ e^{i \omega \Delta\logtime} }{ \left( \omega - \omega_+ \right) \left( \omega - \omega_- \right) } \\
& = \sqrt{2 \pi} ~ e^{i \kappa \etas} ~ e^{ -\frac{3 + \cssq}{2} \Delta\logtime } ~ \times \nonumber \\
& \qquad \qquad \frac{i \cssq \kappa}{\Omega} \sin\left( \Omega \Delta\logtime \right)\text{.}
\end{align}
In both cases, we have performed standard contour integration over the upper half of the complex plane.

The real-space Green's functions are then found by numerically computing the one-dimensional ($\kappa \to \etas$) inverse Fourier transform of the above expressions, and each solution is then a real-space Green's function convolved with the source function $f$:
\begin{align}
\LHe(\logtime, \etas) & = \int_{-\infty}^{+\infty} d\logtime' \int_{-\infty}^{+\infty} d\etas' ~ f(\logtime', \etas') ~ \times \nonumber \\
& \qquad \qquad g_\eps(\logtime', \etas'; \logtime, \etas) \\
& = \int_{-\infty}^{+\infty} d\etas' ~ \epsIC(\etas') ~ g_\eps(0, \etas'; \logtime, \etas) \\
\LHpz(\logtime, \etas) & = \int_{-\infty}^{+\infty} d\etas' ~ \epsIC(\etas') ~ g_p(0, \etas'; \logtime, \etas)\text{.}
\end{align}

\section{Cooper--Frye}
\label{app:CF}

Here we complete our derivation of the surface normal $d^3\Sigma_\mu$ and discuss the dot products $d^3\Sigma_\mu p^\mu$ and $u_\mu p^\mu$ appearing in the Cooper--Frye integration of Sec.~\ref{sec:CF}.

For the cylindrically symmetric hypersurface
\begin{equation}
\Sigma^\mu\left( \tau, \etas \right) = \begin{pmatrix}
\tau \cosh \etas \\
\rFO(\tau, \etas) \\
\tau \sinh \etas
\end{pmatrix}\text{,}
\end{equation}
from which the azimuthal component (erstwhile indexed $2$) has been eliminated, the expression for the surface normal is written
\begin{equation}
d^3\Sigma_\mu = -\epsilon_{\mu\nu\lambda} \frac{\partial \Sigma^\nu}{\partial \tau} \frac{\partial \Sigma^\lambda}{\partial \etas} ~ 2 \pi ~ \rFO ~ d\tau ~ d\etas\text{,} \label{eq:CFdSigma}
\end{equation}
which evaluates to
\begin{align}
d^3\Sigma_0 & = \left[ -\left( \partial_\tau \rFO \right) \tau \cosh \etas + \left( \pdetas \rFO \right) \sinh \etas \right] \nonumber \\
& \qquad \times ~ 2 \pi ~ \rFO ~ d\tau ~ d\etas \\
d^3\Sigma_1 & = 2 \pi ~ \rFO ~ \tau ~ d\tau ~ d\etas \\
d^3\Sigma_3 & = \left[ -\left( \pdetas \rFO \right) \cosh \etas + \left( \partial_\tau \rFO \right) \tau \sinh \etas \right] \nonumber \\
& \qquad \times ~ 2 \pi ~ \rFO ~ d\tau ~ d\etas\text{.}
\end{align}
$\epsilon_{\mu\nu\lambda}$ is the rank-3 Levi--Civita symbol, and $2 \pi ~ \rFO ~ \tau ~ d\tau ~ d\etas$ is the volume element, made annular by integrating out the azimuthal angle.

The \emph{momentum} azimuthal angle $\phi_p$, on the other hand, cannot be eliminated, but as it is an integration variable and we are integrating over all $2\pi$ radians, we can redefine it to be everywhere relative to the outward radial direction.  Consequently only the $\cos\left(\phi_p\right)$ component survives in dot products with the surface normal and flow.

There is nothing to be learned from writing out the terms of the dot products themselves, so we will not do so here.  However, as a final note we underscore that we have returned to Minkowski coordinates, and consequently the dot product $u'_\mu p^\mu$ will use the $(+{}-{}-{}-)$ metric without factors of $\tau$.  ($d\Sigma_\mu$ is defined covariantly, so the metric does not enter when computing $d\Sigma_\mu p^\mu$.)

\bibliography{Trento3D}

\end{document}